\documentclass{mn2e}
\usepackage{psfig}

\def\etal{{\it et al.\ }}
\def\eg{{\it e.g.\ }}

\def\ie{{\it i.e.\ }}

\def\spose#1{\hbox to 0pt{#1\hss}}
\def\approxlt{\mathrel{\spose{\lower 3pt\hbox{$\sim$}}
	\raise 2.0pt\hbox{$<$}}}
\def\approxgt{\mathrel{\spose{\lower 3pt\hbox{$\sim$}}
	\raise 2.0pt\hbox{$>$}}}
\def\approxpropto{\mathrel{\spose{\lower 3pt\hbox{$\sim$}}
	\raise 2.0pt\hbox{$\propto$}}}
\mathchardef\twiddle="2218

\def\multleft#1{\hbox to size{\vbox {\halign {\lft{##}\cr #1}}\hfill}\par}
\def\multright#1{\hbox to size{\vbox {\halign {\rt{##}\cr #1}}\hfill}\par}

\def\today{\ifcase\month\or January\or February\or March\or April\or May\or
      June\or July\or August\or September\or October\or November\or December\fi
      \space\number\day, \number\year}
\def\<{\thinspace}

\def\erg{{\rm\thinspace erg}}

\def\km{{\rm\thinspace km}}

\def\Mpc{{\rm\thinspace Mpc}}

\def\s{{\rm\thinspace s}}

\def\ergps{\hbox{$\erg\s^{-1}\,$}}

\def\kmps{\hbox{$\km\s^{-1}\,$}}

\def\kmpspMpc{\hbox{$\kmps\Mpc^{-1}$}}

\def\OM{\Omega_{\rm m}}
\def\OL{\Omega_{\Lambda}}
\def\OK{\Omega_{\rm k}}

\def\ODE{\Omega_{\rm DE}}

\def\OBH{\Omega_{\rm b}h^2}

\def\fgas{\hbox{$f_{\rm gas}$}}

\def\r2500{\hbox{$r_{2500}$}}
\title[Improved constraints on dark energy from relaxed galaxy clusters]
{Improved constraints on dark energy from Chandra X-ray observations of 
the largest relaxed galaxy clusters}

\author[S.W. Allen et al.]
{\parbox[]{6.in} {S.W. Allen$^1$, D.A. Rapetti$^1$, R.W. Schmidt$^2$, H. Ebeling$^3$, R.G. Morris$^1$ and A.C. Fabian$^4$ \\
\footnotesize
1. Kavli Institute for Particle Astrophysics and Cosmology, Stanford University, 382 Via Pueblo Mall, Stanford, CA 94305-4060, USA.  \\
2. Astronomisches Rechen-Institut, Zentrum f{\"u}r Astronomie der Universit{\"a}t Heidelberg, M{\"o}nchhofstrasse 12-14, 69120 Heidelberg, Germany \\
3. Institute for Astronomy, 2680 Woodlawn Drive, Honolulu, Hawaii 96822, USA \\
4. Institute of Astronomy, Madingley Road, Cambridge CB3 0HA \\
 }}
\begin{document}
\renewcommand{\thefootnote}{\arabic{\footnote}}
\maketitle
\begin{abstract}
We present constraints on the mean matter density, $\OM$, dark energy
density, $\ODE$, and the dark energy equation of state parameter, $w$,
using Chandra measurements of the X-ray gas mass fraction (\fgas) in
42 hot ($kT>5$keV), X-ray luminous, dynamically relaxed galaxy
clusters spanning the redshift range $0.05< z<1.1$.  Using only the
\fgas~data for the six lowest redshift clusters at $z<0.15$, for which
dark energy has a negligible effect on the measurements, we measure
$\OM=0.28\pm0.06$ (68 per cent confidence limits, using standard
priors on the Hubble Constant, $H_0$, and mean baryon density,
$\OBH$). Analyzing the data for all 42 clusters, employing only weak
priors on $H_0$ and $\OBH$, we obtain a similar result on $\OM$ and a
detection of the effects of dark energy on the distances to the
clusters at $\sim 99.99$ per cent confidence, with $\ODE=0.86\pm0.21$
for a non-flat $\Lambda$CDM model. The detection of dark energy is
comparable in significance to recent type Ia supernovae (SNIa) studies
and represents strong, independent evidence for cosmic
acceleration. Systematic scatter remains undetected in the \fgas~data,
despite a weighted mean statistical scatter in the distance
measurements of only $\sim 5$ per cent. For a flat cosmology with a
constant dark energy equation of state, we measure $\OM=0.28\pm0.06$
and $w=-1.14\pm0.31$.  Combining the \fgas~ data with independent
constraints from cosmic microwave background and SNIa studies removes
the need for priors on $\OBH$ and $H_0$ and leads to tighter
constraints: $\OM=0.253\pm0.021$ and $w=-0.98\pm0.07$ for the same
constant$-w$ model. Our most general analysis allows the equation of
state to evolve with redshift. Marginalizing over possible transition
redshifts $0.05<z_{\rm t}<1$, the combined \fgas+CMB+SNIa data set
constrains the dark energy equation of state at late and
early times to be $w_0=-1.05\pm0.29$ and $w_{\rm et}=-0.83\pm0.46$,
respectively, in agreement with the cosmological constant
paradigm. Relaxing the assumption of flatness weakens the constraints
on the equation of state by only a factor $\sim 2$.  Our analysis
includes conservative allowances for systematic uncertainties
associated with instrument calibration, cluster physics, and data
modelling.  The measured small systematic scatter, tight constraint on
$\OM$ and powerful constraints on dark energy from the \fgas~data bode
well for future dark energy studies using the next generation of
powerful X-ray observatories, such as Constellation-X.
\end{abstract}


\section{Introduction}

\noindent The matter content of the largest clusters of galaxies is
expected to provide an almost fair sample of the matter content of the
Universe (\eg White \etal 1993; Eke \etal 1998, Frenk \etal 1999). The
ratio of baryonic-to-total mass in clusters should, therefore, closely
match the ratio of the cosmological parameters $\Omega_{\rm
b}/\Omega_{\rm m}$. The baryonic mass content of clusters is dominated
by the X-ray emitting gas, the mass of which exceeds the mass of
optically luminous material by a factor $\sim 6$, with other sources
of baryonic matter being negligible (Fukugita, Hogan \& Peebles 1998;
Lin \& Mohr 2004). The combination of robust measurements of the
baryonic mass fraction in clusters from X-ray observations together
with a determination of $\Omega_{\rm b}$ from cosmic microwave
background (CMB) data or big-bang nucleosynthesis calculations and a
constraint on the Hubble constant, can therefore be used to measure
$\Omega_{\rm m}$ (\eg White \& Frenk 1991; Fabian 1991; Briel, Henry
\& B\"ohringer 1992; White \etal 1993, David, Jones \& Forman 1995;
White \& Fabian 1995; Evrard 1997; Mohr, Mathiesen \& Evrard 1999;
Ettori \& Fabian 1999; Roussel, Sadat \& Blanchard 2000; Grego \etal
2001; Allen, Schmidt \& Fabian 2002a; Allen \etal 2003, 2004; Ettori,
Tozzi \& Rosati 2003; Sanderson \& Ponman 2003; Lin, Mohr \& Stanford
2003; LaRoque \etal 2006). This method currently provides one of our
best constraints on $\Omega_{\rm m}$ and is remarkably simple and
robust in terms of its underlying assumptions.

Measurements of the apparent evolution of the cluster X-ray gas mass
fraction, hereafter \fgas, can also be used to probe the acceleration
of the Universe (Allen \etal 2004; see also Sasaki 1996, Pen 1997,
Allen \etal 2002a, 2003, Ettori \etal 2003, LaRoque \etal 2006).  This
constraint originates from the dependence of the \fgas{} measurements,
which derive from the observed X-ray gas temperature and density
profiles, on the assumed distances to the clusters, $f_{\rm gas}
\propto d^{1.5}$.\footnote{To understand the origin of the $f_{\rm
gas} \propto d^{1.5}$ dependence, consider a spherical region of
observed angular radius $\theta$ within which the mean gas mass
fraction is measured.  The physical size, $R$, is related to the angle
$\theta$ as $R = \theta d_{\rm A}$.  The X-ray luminosity emitted from
within this region, $L_{\rm X}$, is related to the detected flux,
$F_{\rm X}$, as $L_{\rm X} = 4\pi d_{\rm L}^2 F_{\rm X}$, where
$d_{\rm L}$ is the luminosity distance and $d_{\rm A}=d_{\rm
L}/(1+z)^2$ is the angular diameter distance. Since the X-ray emission
is primarily due to collisional processes (bremsstrahlung and line
emission) and is optically thin, we may also write $L_{\rm X} \propto
n^2 V$, where $n$ is the mean number density of colliding gas
particles and $V$ is the volume of the emitting region, with $V=4\pi
(\theta d_{\rm A})^3/3$.  Considering the cosmological distance
dependences, we see that $n \propto d_{\rm L}/d_{\rm A}^{1.5}$, and
that the observed gas mass within the measurement radius $M_{\rm gas}
\propto nV \propto d_{\rm L}d_{\rm A}^{1.5}$. The total mass, $M_{\rm
tot}$, determined from the X-ray data under the assumption of
hydrostatic equilibrium, $M_{\rm tot} \propto d_{\rm A}$.  Thus, the
X-ray gas mass fraction measured within angle $\theta$ is $f_{\rm
gas}=M_{\rm gas}/M_{\rm tot} \propto d_{\rm L}d_{\rm A}^{0.5}$.} The
expectation from non-radiative hydrodynamical simulations is that for
the largest ($kT\approxgt5$\,keV), dynamically relaxed clusters and
for measurement radii beyond the innermost core ($r \approxgt
r_{2500}$), \fgas{} should be approximately constant with redshift
(Eke \etal 1998; Crain \etal 2007). However, possible systematic
variation of \fgas{} with redshift can be accounted for in a
straightforward manner, so long as the allowed range of such variation
is constrained by numerical simulations or other, complementary data
(Eke \etal 1998; Bialek, Evrard \& Mohr 2001; Muanwong \etal 2002;
Borgani \etal 2004; Kay S. \etal 2004; Ettori \etal 2004, 2006;
Kravtsov, Nagai \& Vikhlinin 2005; Nagai, Vikhlinin \& Kravtsov
2007a).

The first clear detection of cosmic acceleration using the \fgas~
technique was made by Allen \etal (2004) using Chandra observations of
26 hot ($kT\approxgt 5$keV), X-ray luminous ($L_{\rm bol} \approxgt
10^{45}h_{70}^{-2}$\ergps), dynamically relaxed clusters spanning the
redshift range $0.07-0.9$. The total Chandra exposure used in that
work, after all screening procedures were applied, was $\sim 830$ks.
That study led to a $\sim 3\sigma$ detection of the acceleration of
the Universe and a tight constraint on the mean mass density
$\Omega_{\rm m}=0.25\pm0.04$ (see also Allen \etal 2002a, 2003; Ettori
\etal 2003; LaRoque \etal 2006) in excellent agreement with
independent findings from CMB studies (\eg Spergel \etal 2003, 2007),
Type Ia supernovae (SNIa) data (\eg Riess \etal 2004; Astier \etal
2006), galaxy redshift surveys (\eg Cole \etal 2005; Eisenstein \etal
2005; Percival \etal 2007) and X-ray cluster number counts (\eg Mantz
\etal 2007).

Here we present a significant extension of the Allen \etal (2004)
work. Our expanded sample contains 42 clusters spanning the redshift
range $0.05<z<1.1$. We incorporate new, deeper exposures for some of
the original clusters, as well as new targets, approximately doubling
the total exposure time used. Our analysis method incorporates
conservative allowances for systematic uncertainties associated with
instrument calibration, cluster physics and data modelling. As before,
we employ rigorous selection criteria, restricting the analysis to the
hottest, most dynamically relaxed clusters. We show that this leads to
remarkably small intrinsic scatter in the \fgas~measurements, with no
apparent systematic dependence of \fgas~on temperature for clusters
with $kT>5$keV. Our method imposes a minimum of prior constraints and
does not require that the density and temperature profiles of the
X-ray emitting gas follow simple parameterized forms. We make our
\fgas~measurements for each cluster at the radius $r_{2500}$ in the
reference $\Lambda$CDM cosmology, corresponding to an angle
$\theta_{2500}^{\Lambda CDM}$, for which the mean enclosed mass
density is 2500 times the critical density of the Universe at the
redshift of the cluster. This corresponds to about one quarter of the
virial radius\footnote{The virial radius is defined as the radius
within which the density contrast $\Delta_c = 178\, \Omega_{\rm
m}(z)^{0.45}$, with respect to the critical density (Lahav \etal 1991;
Eke \etal 1998).}  and represents a near-optimal choice for Chandra
studies, being sufficiently large to provide small systematic scatter
but not so large as to be hampered by systematic uncertainties in the
background modelling.  We compare our \fgas~measurements to results
from other, independent studies and to the predictions from current
hydrodynamical simulations.

Our analysis of cosmological parameters employs a Markov Chain Monte
Carlo approach, which is efficient and allows for the simple inclusion
of priors and a comprehensive study of the effects of systematic
uncertainties.  We present results based on studies of the \fgas~data
alone (adopting simple priors on $\Omega_{\rm b}h^2$ and $h$) and for
the \fgas~data used in combination with current CMB constraints (in
which case the priors on $\Omega_{\rm b}h^2$ and $h$ can be dropped)
and SNIa data (Astier \etal 2006; Riess \etal 2007; Wood-Vasey \etal
2007; Jha, Riess \& Kirshner 2007). We highlight the power of the data
combinations for cosmological work, particularly in constraining the
mean matter and dark energy densities of the Universe and the dark
energy equation of state.

The \fgas~measurements are quoted for a flat $\Lambda$CDM reference
cosmology with $h= H_0/100$\kmpspMpc=0.7 and $\Omega_{\rm m} =0.3$.

\section{X-ray observations and analysis}

\begin{table*}
\begin{center}
\caption{Summary of the Chandra observations. Columns list the 
target name, observation date, detector used, observation mode, 
net exposure after all cleaning and screening processes were
applied and Right Ascension (R.A.) and Declination (Dec.) for
the X-ray centres. Where multiple observations of a single 
cluster have been used, these are listed separately. 
}\label{table:obs}
\vskip 0 truein
\begin{tabular}{ c c c c c c c c }
&&&&&&&  \\
Name                & ~ &      Date      & Detector & Mode & Exposure (ks) & R.A. (J2000.)  &    DEC. (J2000.)     \\
\hline

Abell 1795(1)       & ~ &   2002 Jun 10  & ACIS-S & VFAINT & 13.2   &   13 48 52.4  & 26 35 38  \\
Abell 1795(2)       & ~ &   2004 Jan 14  & ACIS-S & VFAINT & 14.3   &      ``       &    ``     \\
Abell 1795(3)       & ~ &   2004 Jan 18  & ACIS-I & VFAINT & 9.6    &      ``       &    ``     \\ 
Abell 2029(1)       & ~ &   2000 Apr 12  & ACIS-S & FAINT  & 19.2   &   15 10 56.2  & 05 44 41  \\
Abell 2029(2)       & ~ &   2004 Jan 08  & ACIS-S & FAINT  & 74.8   &      ``       &    ``     \\
Abell 2029(3)       & ~ &   2004 Dec 17  & ACIS-I & VFAINT & 9.4    &      ``       &    ``     \\
Abell 478(1)        & ~ &   2001 Jan 27  & ACIS-S & FAINT  & 39.9   &   04 13 25.2  & 10 27 55  \\
Abell 478(2)        & ~ &   2004 Sep 13  & ACIS-I & VFAINT & 7.4    &      ``       &    ``     \\
PKS0745-191(1)      & ~ &   2001 Jun 16  & ACIS-S & VFAINT & 17.4   &   07 47 31.7  &-19 17 45  \\
PKS0745-191(2)      & ~ &   2004 Sep 24  & ACIS-I & VFAINT & 9.2    &      ``       &    ``     \\
Abell 1413          & ~ &   2001 May 16  & ACIS-I & VFAINT & 64.5   &   11 55 18.1  & 23 24 17  \\
Abell 2204(1)       & ~ &   2000 Jul 29  & ACIS-S &  FAINT & 10.1   &   16 32 47.2  & 05 34 32  \\
Abell 2204(2)       & ~ &   2004 Sep 20  & ACIS-I & VFAINT & 8.5    &      ``       &    ``     \\
Abell 383(1)        & ~ &   2000 Nov 16  & ACIS-S &  FAINT & 18.0   &   02 48 03.5  &-03 31 45  \\
Abell 383(2)        & ~ &   2000 Nov 16  & ACIS-I & VFAINT & 17.2   &      ``       &    ``     \\
Abell 963           & ~ &   2000 Oct 11  & ACIS-S & FAINT  & 35.8   &   10 17 03.8  & 39 02 49  \\
RXJ0439.0+0520      & ~ &   2000 Aug 29  & ACIS-I & VFAINT & 7.6    &   04 39 02.3  & 05 20 44  \\
RXJ1504.1-0248      & ~ &   2005 Mar 20  & ACIS-I & VFAINT & 29.4   &   15 04 07.9  &-02 48 16  \\
Abell 2390          & ~ &   2003 Sep 11  & ACIS-S & VFAINT & 79.2   &   21 53 36.8  & 17 41 44  \\
RXJ2129.6+0005      & ~ &   2000 Oct 21  & ACIS-I & VFAINT & 7.6    &   21 29 39.9  & 00 05 20  \\
Abell 1835(1)       & ~ &   1999 Dec 11  & ACIS-S & FAINT  & 18.0   &   14 01 01.9  & 02 52 43  \\
Abell 1835(2)       & ~ &   2000 Apr 29  & ACIS-S & FAINT  & 10.3   &      ``       &    ``     \\
Abell 611           & ~ &   2001 Nov 03  & ACIS-S & VFAINT & 34.5   &   08 00 56.8  & 36 03 24  \\
Zwicky 3146         & ~ &   2000 May 10  & ACIS-I &  FAINT & 41.4   &   10 23 39.4  & 04 11 14  \\
Abell 2537          & ~ &   2004 Sep 09  & ACIS-S & VFAINT & 36.0   &   23 08 22.1  &-02 11 29  \\
MS2137.3-2353(1)    & ~ &   1999 Nov 18  & ACIS-S & VFAINT & 20.5   &   21 40 15.2  &-23 39 40  \\
MS2137.3-2353(2)    & ~ &   2003 Nov 18  & ACIS-S & VFAINT & 26.6   &       ``      &    ``     \\
MACSJ0242.6-2132    & ~ &   2002 Feb 07  & ACIS-I & VFAINT & 10.2   &   02 42 35.9  &-21 32 26  \\
MACSJ1427.6-2521    & ~ &   2002 Jun 29  & ACIS-I & VFAINT & 14.7   &   14 27 39.4  &-25 21 02  \\
MACSJ2229.8-2756    & ~ &   2002 Nov 13  & ACIS-I & VFAINT & 11.8   &   22 29 45.3  &-27 55 37  \\
MACSJ0947.2+7623    & ~ &   2000 Oct 20  & ACIS-I & VFAINT & 9.6    &   09 47 13.1  & 76 23 14  \\
MACSJ1931.8-2635    & ~ &   2002 Oct 20  & ACIS-I & VFAINT & 12.2   &   19 31 49.6  &-26 34 34  \\
MACSJ1115.8+0129    & ~ &   2003 Jan 23  & ACIS-I & VFAINT & 10.2   &   11 15 52.1  & 01 29 53  \\
&&&&&&& \\             
\hline                      
\end{tabular}
\end{center}
\end{table*}

\addtocounter{table}{-1}
\begin{table*}
\begin{center}
\caption{Summary of the Chandra observations -- continued }
\vskip 0 truein
\begin{tabular}{ c c c c c c c c }
&&&&&&&  \\
Name                & ~ &      Date      & Detector & Mode & Exposure (ks) & R.A. (J2000.)  &    DEC. (J2000.)     \\
\hline

MACSJ1532.9+3021(1) & ~ &   2001 Aug 26  & ACIS-S & VFAINT & 9.4    &   15 32 53.9  & 30 20 59  \\
MACSJ1532.9+3021(2) & ~ &   2001 Sep 06  & ACIS-I & VFAINT & 9.2    &       ``      &    ``     \\
MACSJ0011.7-1523(1) & ~ &   2002 Nov 20  & ACIS-I & VFAINT & 18.2   &   00 11 42.9  &-15 23 22  \\
MACSJ0011.7-1523(2) & ~ &   2005 Jun 28  & ACIS-I & VFAINT & 32.1   &       ``      &    ``     \\
MACSJ1720.3+3536(1) & ~ &   2002 Nov 03  & ACIS-I & VFAINT & 16.6   &   17 20 16.8  & 35 36 27  \\
MACSJ1720.3+3536(2) & ~ &   2005 Nov 22  & ACIS-I & VFAINT & 24.8   &       ``      &    ``     \\
MACSJ0429.6-0253    & ~ &   2002 Feb 07  & ACIS-I & VFAINT & 19.1   &   04 29 36.1  &-02 53 08  \\
MACSJ0159.8-0849(1) & ~ &   2002 Oct 02  & ACIS-I & VFAINT & 14.1   &   01 59 49.4  &-08 49 58  \\
MACSJ0159.8-0849(2) & ~ &   2004 Dec 04  & ACIS-I & VFAINT & 28.9   &       ``      &    ``     \\
MACSJ2046.0-3430    & ~ &   2005 Jun 28  & ACIS-I & VFAINT & 8.9    &   20 46 00.5  &-34 30 17  \\
MACSJ1359.2-1929    & ~ &   2005 Mar 17  & ACIS-I & VFAINT & 9.2    &   13 59 10.3  &-19 29 24  \\
MACSJ0329.7-0212(1) & ~ &   2002 Dec 24  & ACIS-I & VFAINT & 16.8   &   03 29 41.7  &-02 11 48  \\
MACSJ0329.7-0212(2) & ~ &   2004 Dec 06  & ACIS-I & VFAINT & 31.1   &       ``      &    ``     \\
RXJ1347.5-1145(1)   & ~ &   2000 Mar 03  & ACIS-S & VFAINT & 8.6    &   13 47 30.6  &-11 45 10  \\
RXJ1347.5-1145(2)   & ~ &   2000 Apr 29  & ACIS-S & FAINT  & 10.0   &       ``      &    ``     \\
RXJ1347.5-1145(3)   & ~ &   2003 Sep 03  & ACIS-I & VFAINT & 49.3   &       ``      &    ``     \\
3C295(1)            & ~ &   1999 Aug 30  & ACIS-S & FAINT  & 15.4   &   14 11 20.5  & 52 12 10  \\
3C295(2)            & ~ &   2001 May 18  & ACIS-I & FAINT  & 72.4   &       ``      &    ``     \\
MACSJ1621.6+3810(1) & ~ &   2002 Oct 18  & ACIS-I & VFAINT & 7.9    &   16 21 24.8  & 38 10 09  \\
MACSJ1621.6+3810(2) & ~ &   2004 Dec 11  & ACIS-I & VFAINT & 32.2   &       ``      &    ``     \\
MACSJ1621.6+3810(3) & ~ &   2004 Dec 25  & ACIS-I & VFAINT & 26.1   &       ``      &    ``     \\
MACS1427.3+4408     & ~ &   2005 Feb 12  & ACIS-I & VFAINT & 8.70   &   14 27 16.2  & 44 07 31  \\
MACSJ1311.0-0311    & ~ &   2005 Apr 20  & ACIS-I & VFAINT & 56.2   &   13 11 01.6  &-03 10 40  \\
MACSJ1423.8+2404    & ~ &   2003 Aug 18  & ACIS-S & VFAINT & 113.5  &   14 23 47.9  & 24 04 43  \\
MACSJ0744.9+3927(1) & ~ &   2001 Nov 12  & ACIS-I & VFAINT & 17.1   &   07 44 52.9  & 39 27 27  \\
MACSJ0744.9+3927(2) & ~ &   2003 Jan 04  & ACIS-I & VFAINT & 15.6   &       ``      &    ``     \\
MACSJ0744.9+3927(3) & ~ &   2004 Dec 03  & ACIS-I & VFAINT & 41.3   &       ``      &    ``     \\
MS1137.5+6625       & ~ &   1999 Sep 30  & ACIS-I & VFAINT & 103.8  &   11 40 22.4  & 66 08 15  \\
ClJ1226.9+3332(1)   & ~ &   2003 Jan 27  & ACIS-I & VFAINT & 25.7   &   12 26 58.1  & 33 32 47  \\
ClJ1226.9+3332(2)   & ~ &   2004 Aug 07  & ACIS-I & VFAINT & 26.3   &       ``      &    ``     \\
CL1415.2+3612       & ~ &   2003 Sep 16  & ACIS-I & VFAINT & 75.1   &   14 15 11.2  & 36 12 02  \\
3C186               & ~ &   2002 May 16  & ACIS-S & VFAINT & 15.4   &   07 44 17.5  & 37 53 17  \\
&&&&&&& \\             
\hline                      
\end{tabular}
\end{center}
\end{table*}

\subsection{Sample selection}
\label{section:selection}

Our sample consists of 42 hot, X-ray luminous, dynamically relaxed
galaxy clusters spanning the redshift range $0.05<z<1.1$.  The systems
have mass weighted X-ray temperatures measured within $r_{2500}$,
$kT_{2500}\approxgt5$\,keV and exhibit a high degree of dynamical
relaxation in their Chandra images (Million \etal 2007, in
prep.), with sharp central X-ray surface brightness peaks,
short central cooling times ($t_{\rm cool}\leq$ a few $10^9$\,yr)
minimal isophote centroid variations (\eg Mohr \etal 1995) and low
X-ray power ratios (Buote \& Tsai 1995, 1996; Jeltema \etal
2005). Although target selection is based only on these morphological
X-ray characteristics, the clusters also exhibit other signatures of
dynamical relaxation including minimal evidence for departures from
hydrostatic equilibrium in X-ray pressure maps (Million \etal 2007, in
prep.). The notable exceptions are Abell 2390, RXJ1347.5-1145,
MACS1427.3+4408 and MACSJ0744.9+3927, for which clear substructure is
observed between position angles of 255-15 degrees, 90-190 degrees,
160-280 degrees and 210-330 degrees, respectively (Allen, Schmidt \&
Fabian 2002b; Morris \etal 2007, in prep.; Million \etal 2007, in
prep.). The regions associated with obvious substructure in these
clusters have been excluded from the analysis.  The bulk of the
clusters at $z>0.3$ were identified in the MACS survey (Ebeling, Edge
\& Henry 2001; Ebeling \etal 2007).  Of the 70 MACS clusters with
sufficient data on the Chandra archive at the time of writing to
enable detailed spatially-resolved spectroscopy, 22/70 are identified
as being sufficiently relaxed to be included in the present study.

The restriction to clusters with the highest possible degree of
dynamical relaxation, for which the assumption of hydrostatic
equilibrium should be most valid, minimizes systematic scatter in the
\fgas~data (Section~\ref{section:scatter}) and allows for the most
precise and robust determination of cosmological parameters. The
restriction to the $hottest$ ($kT>5$keV), relaxed systems further
simplifies the analysis: for galaxies, groups and clusters with
$kT\approxlt4$keV, the baryonic mass fraction is both expected and
observed to rise systematically with increasing temperature, with the
systematic scatter being largest in the coolest systems (\eg Bialek
\etal 2001; Muanwong \etal 2002; Ettori \etal 2004; Kravtsov,
\etal  2005; Vikhlinin \etal 2006).  As shown in
Sections~\ref{section:fgasres2} and~\ref{section:scatter}, for the hot,
relaxed clusters studied here, \fgas{} exhibits no dependence on
temperature and the intrinsic scatter is small.

\subsection{Data reduction}

The Chandra observations were carried out using the Advanced CCD
Imaging Spectrometer (ACIS) between 1999 August 30 and 2005 June 28.
The standard level-1 event lists produced by the Chandra pipeline
processing were reprocessed using the $CIAO$ (version 3.2.2) software
package, including the appropriate gain maps and calibration products. Bad
pixels were removed and standard grade selections applied. Where
possible, the extra information available in VFAINT mode was used to
improve the rejection of cosmic ray events. The data were cleaned to
remove periods of anomalously high background using the standard
energy ranges and time bins recommended by the Chandra X-ray Center.
The net exposure times after cleaning are summarized in
Table~\ref{table:obs}. The total good exposure is 1.63 Ms,
approximately twice that of the Allen \etal (2004) study.

\subsection{Spectral analysis}
\label{section:spectra}

The spectral analysis was carried out using an updated version of the
techniques described by Allen \etal (2004) and Schmidt \& Allen
(2007).  In brief, concentric annular spectra were extracted from the
cleaned event lists, centred on the coordinates listed in
Table~\ref{table:obs}. Emission associated with X-ray point sources or
obvious substructure (Table~\ref{table:exclude}) was excluded.  The
spectra were analysed using XSPEC (version 11.3; Arnaud 1996), the
MEKAL plasma emission code (Kaastra \& Mewe 1993; incorporating the
Fe-L calculations of Liedhal, Osterheld \& Goldstein 1995) and the
photoelectric absorption models of Balucinska-Church \& McCammon
(1992).  The emission from each spherical shell was modelled as a
single phase plasma. The abundances of the elements in each shell
were assumed to vary with a common ratio, $Z$, with respect to Solar
values. The absorbing column densities were fixed to the Galactic values
determined from HI studies (Dickey \& Lockman 1990), with the
exception of Abell 478 and PKS0745-191 where the value was allowed to
fit freely. (For Abell 478, the absorbing column density was allowed
to vary as a function of radius, as was shown to be required
by Allen \etal 1993).  We have included
standard correction factors to account for time-dependent
contamination along the instrument light path.  In addition, we have
incorporated a small correction to the High Resolution Mirror Assembly
model in CIAO 3.2.2, which takes the form of an `inverse' edge with an
energy, E=2.08\,keV and optical depth $\tau=-0.1$ (H. Marshall,
private communication) and also boosted the overall effective area by six
per cent, to better match later calibration data (A. Vikhlinin,
private communication). These corrections lead to an excellent match
with results based on later calibration data, available in CIAO 3.4.  
Only data in the $0.8-7.0$ keV energy
range were used in the analysis (with the exceptions of the earliest
observations of 3C 295, Abell 1835 and Abell 2029, where a wider 0.6
to 7.0 keV band was used to enable better modelling of the
soft X-ray background).

For the nearer clusters ($z<0.3$), background spectra were extracted
from the blank-field data sets available from the Chandra X-ray
Center. These were cleaned in an identical manner to the target
observations. In each case, the normalizations of the background files
were scaled to match the count rates in the target observations
measured in the 9.5-12keV band. Where required, \eg due to the
presence of strong excess soft emission in the field, a spectral model
for additional soft background emission was included in the
analysis. For the more distant systems (as well as for the first
observation of Abell 1835, the ACIS-I observation of Abell 383, and
the observations of Abell 2537, RXJ 2129.6+0005 and Zwicky 3146)
background spectra were extracted from appropriate, source free
regions of the target data sets. (We have confirmed that similar
results are obtained using the blank-field background data sets.) In
order to minimize systematic errors, we have restricted our
spectral analysis to radii within which systematic uncertainties in
the background subtraction (established by the comparison of different
background subtraction methods) are smaller than the statistical
uncertainties in the results. All results are drawn from ACIS chips
0,1,2,3 and 7 which have the most accurate calibration, although ACIS
chip 5 was also used to study the soft X-ray background in ACIS-S
observations.

Separate photon-weighted response matrices and effective area files
were constructed for each region using calibration files appropriate
for the period of observations. The spectra for all annuli for a given
cluster were modelled simultaneously in order to determine the
deprojected X-ray gas temperature and metallicity profiles, under the
assumption of spherical symmetry. The extended C-statistic, 
available in XSPEC, was used for all spectral fitting.

\subsection{Measuring the mass profiles}

The details of the mass analysis and results on the total mass and
dark matter profiles are presented by Schmidt \& Allen (2007).  In
brief, X-ray surface brightness profiles in the 0.8-7.0keV band were
extracted from background subtracted, flat-fielded Chandra images with
$0.984\times0.984$arcsec$^2$ pixels. The profiles were centered on the
coordinates listed in Table~\ref{table:obs}.  Under the assumptions of
hydrostatic equilibrium and spherical symmetry, the observed X-ray
surface brightness profiles and deprojected X-ray gas temperature
profiles may together be used to determine the X-ray emitting gas mass
and total mass profiles in the clusters. For this analysis, we have
used an enhanced version of the Cambridge X-ray deprojection code
described by \eg White, Jones \& Forman (1997). This method is
particularly well suited to the present task in that it does not use
parametric fitting functions for the X-ray temperature, gas density or
surface brightness in measuring the mass; the use of such functions
introduces strong priors that complicate the interpretation of results
and, in particular, can lead to an underestimation of
uncertainties. The only additional important assumption in the
analysis is the choice of a Navarro, Frenk \& White (1995, 1997;
hereafter NFW) model to parameterize the total (luminous-plus-dark)
mass distributions:

\vspace{0.1cm}
\begin{equation} \rho(r) = {{\rho_{\rm c}(z) \delta_{\rm c}} \over {
({r/r_{\rm s}}) \left(1+{r/r_{\rm s}} \right)^2}},
\end{equation}
\vspace{0.1cm}

\noindent where $\rho(r)$ is the mass density, $\rho_{\rm c}(z) =
3H(z)^2/ 8 \pi G$ is the critical density for closure at redshift $z$,
$r_{\rm s}$ is the scale radius, $c$ is the concentration parameter
(with $c=r_{200}/r_{\rm s}$) and $\delta_{\rm c} = {200 c^3 / 3 \left[
{{\rm ln}(1+c)-{c/(1+c)}}\right]}$.\footnote{Note that the outermost
pressure, at the limit of the X-ray surface brightness profile, is
fixed using an iterative method that ensures a smooth, power law
pressure gradient in these regions. The model temperature profiles,
for radii spanned by the spectral data, are not sensitive to any
reasonable choices for the outer pressures.}  Schmidt \& Allen (2007)
show that the NFW model provides a good description of the mass
distributions in the clusters studied here.

Given the observed surface brightness profile and a particular choice
of parameters for the total mass profile, the deprojection code is
used to predict the temperature profile of the X-ray gas. (In detail,
the median model temperature profile determined from 100 Monte-Carlo
simulations for each mass model is used.) This model temperature
profile is then compared with the observed spectral, deprojected
temperature profile and the goodness of fit is calculated using the
sum over all temperature bins

\vspace{0.1cm}
\begin{equation} \chi^2 = \sum_{\,\rm all\,bins}\,\left(
\frac{T_{\,\rm obs} - T_{\,\rm model}}{\sigma_{\,\rm obs}} \right)^2,
\end{equation} 
\vspace{0.1cm}

\noindent where $T_{\,\rm obs}$ is the observed, spectral
deprojected temperature profile and $T_{\,\rm model}$ is the model,
rebinned to the same spatial scale. For each cluster, the mass
parameters are stepped over a grid of values and the best-fit
values and uncertainties determined via $\chi^2$ minimization
techniques. The X-ray emitting gas density, pressure, entropy, cooling
time and mass, and the integrated X-ray gas mass fraction, $f_{\rm
gas}$, are then determined in a straightforward manner from the
Monte-Carlo simulations and $\chi^2$ values at each grid point.

A number of systematic issues affect the accuracy of the $f_{\rm gas}$
measurements and their interpretation; these are discussed in detail
in Section~\ref{section:modelling}. In particular, our analysis
incorporates allowances for effects associated with calibration and
modelling uncertainties and non-thermal pressure support in the X-ray
emitting gas, employing priors that span conservative ranges for the
likely magnitudes of these effects.

Finally, for a number of the clusters, noticeable substructure is
present at small radii. This is likely to result from interactions
between the central radio sources and surrounding gas (\eg B\"ohringer
\etal 1993; Fabian \etal 2000, 2003a, 2005, 2006; Birzan \etal 2004; Dunn \&
Fabian 2004; Forman \etal 2005; Dunn, Fabian \& Taylor 2005;
Allen \etal 2006; Rafferty \etal 2006) and/or `sloshing' of the X-ray
emitting gas within the central potentials (\eg Churazov \etal 2003; 
Markevitch \etal 2003; Ascasibar \& Markevitch
2006). The regions affected by such substructure are listed in
Table~\ref{table:exclude}. A systematic uncertainty of $\pm 30$ per
cent has been added in quadrature to all spectral results determined
from these regions, leading to them having little weight in the mass
analysis.

\begin{table}
\begin{center}
\caption{Clusters with regions of localized substructure 
that have been excluded or down-weighted in the analysis. 
Column two lists the position angles (PA) 
that have been excluded in the case of 
Abell 2390, RXJ1347.5-1145, MACS1427.3+4408 and MACSJ0744.9+3927. 
Column 3 lists the radii (in $h_{70}^{-1}$kpc) 
within which the spectral data have been down-weighted 
by including a systematic uncertainty of 
$\pm30$ per cent in quadrature with the statistical
errors on the temperature
measurements.}\label{table:exclude}
\vskip 0 truein
\begin{tabular}{ c c c c }
Cluster             & &  Excluded P.A.& Down-weighted r \\
\hline
Abell 1795          & &    --        &     75  \\
Abell 2029          & &    --        &     30  \\
Abell 478           & &    --        &     15  \\
PKS0745-191         & &    --        &     55  \\
Abell 1413          & &    --        &     40  \\
Abell 2204          & &    --        &     75  \\
Abell 383           & &    --        &     40  \\
RXJ1504.1-0248      & &    --        &     80  \\
Abell 2390          & &  $255-15$    &     50  \\
RXJ2129.6+0005      & &     --       &     40  \\
Zwicky 3146         & &    --        &     240 \\
Abell 2537          & &    --        &     40  \\
MACSJ2229.8-2756    & &    --        &     40  \\
MACSJ0947.2+7623    & &    --        &     40  \\
MACSJ1931.8-2635    & &    --        &     40  \\
MACSJ1115.8+0129    & &    --        &     85  \\
MACSJ1532.9+3021    & &    --        &     40  \\
RXJ1347.5-1145      & &  $90-190$    &     --  \\
MACSJ1621.6+3810    & &    --        &     45  \\
MACSJ1427.3+4408    & &  $160-280$   &     --  \\
MACSJ0744.9+3927    & &  $210-330$   &     --  \\
&&& \\             
\hline                      
\end{tabular}
\end{center}
\end{table}

\subsection{The stellar baryonic mass fraction}
\label{section:opt}

Observations of nearby and intermediate redshift clusters show that
for clusters in the mass/temperature range studied here, the average
mass fraction in stars (in galaxies and intracluster light
combined) $f_{\rm star} \sim 0.16h_{70}^{0.5} f_{\rm gas}$ (Lin \&
Mohr 2004; see also White \etal 1993; Fukugita, Hogan \& Peebles 1998;
Balogh \etal 2001).

For the present analysis, we ideally require the ratio $s=f_{\rm
star}/f_{\rm gas}$ measured within $r_{2500}$ for each cluster.
However, such measurements are not yet available for the bulk
of the clusters studied here. For hot, massive clusters, the relative
contribution of the central dominant galaxy to the overall cluster
light is less significant than for cooler, less massive systems (\eg
Lin \& Mohr 2004). We have therefore assumed that the stellar mass
fraction within $r_{2500}$ is similar to that measured within the
virial radius \ie $s= 0.16h_{70}^{0.5}$, but have both included a
conservative 30 per cent Gaussian uncertainty in this value and
allowed for evolution at the $\pm 20$ per cent level, per unit
redshift interval. Since the stellar mass accounts for only $\sim 14$
per cent of the overall baryon budget within $r_{2500}$ and less than
2 per cent of the total mass, these systematic uncertainties do not
have a large effect on the overall error budget.  A program to measure
the evolution of the optical baryonic mass content of the largest
relaxed clusters is underway.

\section{The X-ray gas mass fraction measurements}
\label{section:fgasres}

\begin{figure*}
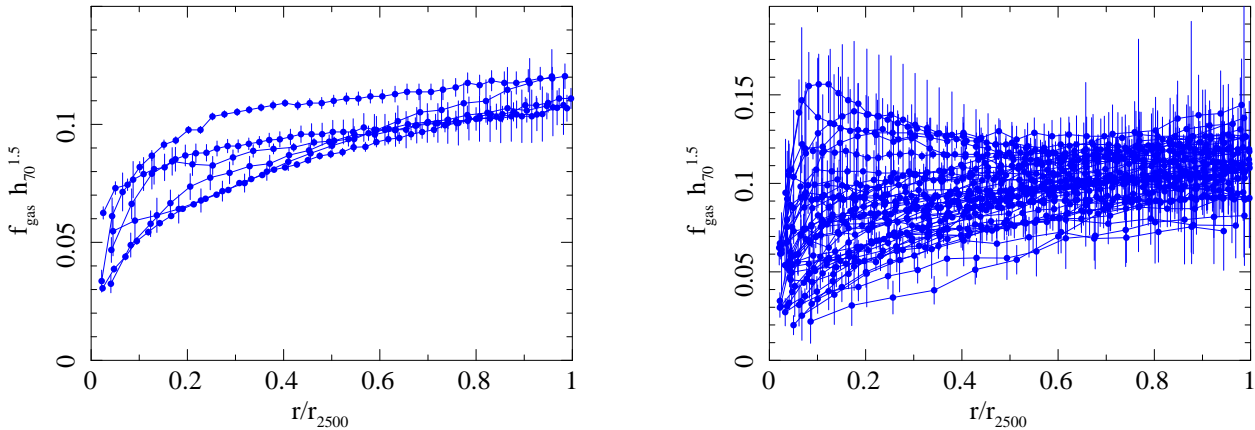

\vspace{0.2cm}
\hbox{
\hspace{0.2cm}\psfig{figure=./fgas_r_lowz_single.ps,width=0.43 \textwidth,angle=270}
\hspace{1.3cm}\psfig{figure=./fgas_r_lcdm_single.ps,width=0.43 \textwidth,angle=270}
}
\caption{The X-ray gas mass fraction profiles for the $\Lambda$CDM reference 
cosmology ($\Omega_{\rm m}=0.3$, 
$\Omega_{\Lambda}=0.7$, $h=0.7$) with the radial axes scaled in units of 
$r_{2500}$. (a: left panel) Results for the 
six lowest redshift clusters with $z \approxlt 0.15$ (b: right panel) 
Results for the entire sample. Note $f_{\rm gas}(r)$
is an integrated quantity and so error bars on neighbouring points
in a profile are correlated.}\label{fig:fgasr}
\end{figure*}

\begin{figure*}
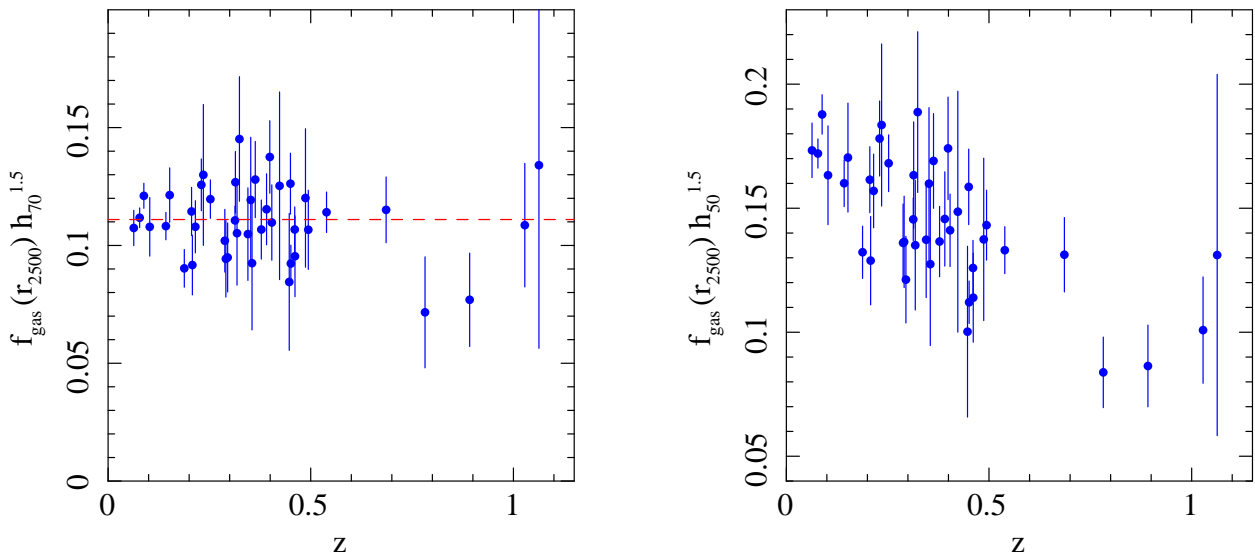

\vspace{0.2cm}
\hbox{
\hspace{0.2cm}\psfig{figure=./fgas_z_lcdm.ps,width=0.43 \textwidth,angle=270}
\hspace{1.3cm}\psfig{figure=./fgas_z_scdm.ps,width=0.43 \textwidth,angle=270}
}
\caption{The apparent variation of the X-ray gas mass fraction
measured within $r_{2500}$ as a function of redshift for the (a: left
panel) reference $\Lambda$CDM and (b: right panel) reference SCDM
($\Omega_{\rm m}=1.0$, $\Omega_{\Lambda}=0.0$, $h=0.5$) cosmologies.
The plotted error bars are statistical root-mean-square $1\sigma$
uncertainties. The global, absolute normalization of the $f_{\rm gas}$
values should be regarded as uncertain at the $\sim 10-15$ per cent
level due to systematic uncertainties in instrument calibration,
modelling and the level of non-thermal pressure support (Section
4.2).}\label{fig:fgasz}
\end{figure*}

\subsection{New \fgas~measurements}
\label{section:fgasres2}

\begin{table*}
\begin{center}
\caption{The redshifts, $r_{2500}$ values, mean mass-weighted
temperatures within $r_{2500}$, and the X-ray gas mass fractions
within $r_{2500}$ for the reference $\Lambda$CDM cosmology. Error bars
are statistical uncertainties and are quoted at the 68 per cent
confidence level. A systematic uncertainty of $\sim 10-15$ per cent is
associated with the global, absolute normalization of the $f_{\rm
gas}$ values due to uncertainties in instrument calibration, X-ray
modelling and non-thermal pressure support (Section 4.2). The
redshifts for the MACS clusters are from Ebeling \etal (2007, in
preparation).  }\label{table:fgas}
\vskip -0.1truein
\begin{tabular}{ c c c c c c }
&&&&&  \\
\multicolumn{1}{c}{} &
\multicolumn{1}{c}{} &
\multicolumn{1}{c}{z} &
\multicolumn{1}{c}{{$r_{2500}\,(h_{70}^{-1}$}kpc)} &
\multicolumn{1}{c}{{$kT_{2500}$ }} &
\multicolumn{1}{c}{{$f_{\rm gas}\, h_{70}^{1.5}$}} \\
\hline                                                                      
Abell 1795             & &  0.063 &  $570_{-24}^{+18}$   & $6.51\pm0.23 $   & $0.1074\pm0.0075$  \\
Abell 2029             & &  0.078 &  $611_{-13}^{+10}$   & $8.58\pm0.44 $   & $0.1117\pm0.0042$  \\
Abell 478              & &  0.088 &  $643_{-15}^{+16}$   & $7.99\pm0.43 $   & $0.1211\pm0.0053$  \\
PKS0745-191            & &  0.103 &  $682_{-41}^{+42}$   & $9.50\pm1.13 $   & $0.1079\pm0.0124$  \\
Abell 1413             & &  0.143 &  $599_{-19}^{+17}$   & $7.80\pm0.35 $   & $0.1082\pm0.0058$  \\
Abell 2204             & &  0.152 &  $628_{-24}^{+38}$   & $10.51\pm2.54$   & $0.1213\pm0.0116$  \\
Abell 383              & &  0.188 &  $502_{-23}^{+25}$   & $5.36\pm0.23 $   & $0.0903\pm0.0080$  \\
Abell 963              & &  0.206 &  $540_{-27}^{+24}$   & $7.26\pm0.28 $   & $0.1144\pm0.0102$  \\
RXJ0439.0+0521         & &  0.208 &  $454_{-25}^{+37}$   & $4.86\pm0.45 $   & $0.0917\pm0.0127$  \\
RXJ1504.1-0248         & &  0.215 &  $671_{-33}^{+44}$   & $9.32\pm0.59 $   & $0.1079\pm0.0111$  \\
Abell 2390             & &  0.230 &  $662_{-30}^{+42}$   & $11.72\pm1.43$   & $0.1257\pm0.0110$  \\
RXJ2129.6+0005         & &  0.235 &  $507_{-57}^{+65}$   & $7.38\pm0.88 $   & $0.1299\pm0.0299$  \\
Abell 1835             & &  0.252 &  $684_{-26}^{+27}$   & $10.57\pm0.62$   & $0.1197\pm0.0082$  \\
Abell 611              & &  0.288 &  $518_{-30}^{+43}$   & $7.39\pm0.48 $   & $0.1020\pm0.0133$  \\
Zwicky 3146            & &  0.291 &  $679_{-66}^{+66}$   & $8.27\pm1.08 $   & $0.0943\pm0.0163$  \\
Abell 2537             & &  0.295 &  $518_{-33}^{+57}$   & $8.12\pm0.78 $   & $0.0949\pm0.0147$  \\
MS2137.3-2353          & &  0.313 &  $479_{-10}^{+18}$   & $5.65\pm0.30 $   & $0.1106\pm0.0061$  \\
MACSJ0242.6-2132       & &  0.314 &  $478_{-20}^{+29}$   & $5.51\pm0.47 $   & $0.1268\pm0.0131$  \\
MACSJ1427.6-2521       & &  0.318 &  $412_{-37}^{+42}$   & $5.24\pm0.77 $   & $0.1052\pm0.0220$  \\
MACSJ2229.8-2756       & &  0.324 &  $414_{-29}^{+41}$   & $5.42\pm0.68 $   & $0.1452\pm0.0265$  \\
MACSJ0947.2+7623       & &  0.345 &  $594_{-49}^{+65}$   & $7.80\pm0.69 $   & $0.1048\pm0.0196$  \\
MACSJ1931.8-2635       & &  0.352 &  $581_{-46}^{+131}$  & $7.49\pm0.77 $   & $0.1193\pm0.0266$  \\
MACSJ1115.8+0129       & &  0.355 &  $664_{-108}^{+118}$ & $8.92\pm1.31 $   & $0.0925\pm0.0283$  \\
MACSJ1532.9+3021       & &  0.363 &  $543_{-33}^{+45}$   & $7.69\pm1.34 $   & $0.1280\pm0.0162$  \\
MACSJ0011.7-1523       & &  0.378 &  $497_{-27}^{+40}$   & $6.56\pm0.37 $   & $0.1067\pm0.0125$  \\
MACSJ1720.3+3536       & &  0.391 &  $520_{-32}^{+39}$   & $8.11\pm0.55 $   & $0.1153\pm0.0151$  \\
MACSJ0429.6-0253       & &  0.399 &  $439_{-24}^{+19}$   & $6.10\pm0.58 $   & $0.1375\pm0.0154$  \\
MACSJ0159.8-0849       & &  0.404 &  $597_{-48}^{+33}$   & $10.62\pm0.69$   & $0.1097\pm0.0160$  \\
MACSJ2046.0-3430       & &  0.423 &  $413_{-50}^{+62}$   & $5.81\pm1.02 $   & $0.1253\pm0.0398$  \\
MACSJ1359.2-1929       & &  0.447 &  $458_{-56}^{+91}$   & $6.73\pm0.96 $   & $0.0845\pm0.0290$  \\
MACSJ0329.7-0212       & &  0.450 &  $481_{-23}^{+26}$   & $6.85\pm0.45 $   & $0.1262\pm0.0129$  \\
RXJ1347.5-1144         & &  0.451 &  $776_{-31}^{+43}$   & $14.54\pm1.08$   & $0.0923\pm0.0078$  \\
3C295                  & &  0.461 &  $419_{-15}^{+20}$   & $5.09\pm0.42 $   & $0.1067\pm0.0096$  \\
MACSJ1621.6+3810       & &  0.461 &  $496_{-39}^{+53}$   & $9.15\pm1.01 $   & $0.0954\pm0.0172$  \\
MACS1427.3+4408        & &  0.487 &  $428_{-36}^{+67}$   & $6.65\pm1.40 $   & $0.1201\pm0.0294$  \\
MACSJ1311.0-0311       & &  0.494 &  $461_{-26}^{+30}$   & $6.07\pm0.71 $   & $0.1066\pm0.0168$  \\
MACSJ1423.8+2404       & &  0.539 &  $467_{-14}^{+18}$   & $7.80\pm0.44 $   & $0.1141\pm0.0086$  \\
MACSJ0744.9+3927       & &  0.686 &  $466_{-23}^{+40}$   & $8.67\pm0.98 $   & $0.1151\pm0.0140$  \\
MS1137.5+6625          & &  0.782 &  $435_{-44}^{+84}$   & $6.89\pm0.78 $   & $0.0716\pm0.0235$  \\
ClJ1226.9+3332         & &  0.892 &  $521_{-54}^{+123}$  & $11.95\pm1.97$   & $0.0769\pm0.0198$  \\
CL1415.2+3612          & &  1.028 &  $278_{-25}^{+33}$   & $5.59\pm0.84 $   & $0.1086\pm0.0262$  \\
3C186                  & &  1.063 &  $292_{-57}^{+54}$   & $5.62\pm1.00 $   & $0.1340\pm0.0777$  \\
&&&&& \\                                                                             
\hline                                                                                                                                       
\end{tabular}                                                                                                                                
\end{center}
\end{table*}

As mentioned above, in compiling the results on the X-ray gas mass
fraction, $f_{\rm gas}$, we have adopted a canonical measurement
radius of $r_{2500}$. The $r_{2500}$ value for each cluster is
determined directly from the Chandra data, with confidence limits
calculated from the $\chi^2$ grids. In general, the values are
well-matched to the outermost radii at which reliable temperature
measurements can be made from the Chandra data, given systematic
uncertainties associated with the background modelling.

Fig. \ref{fig:fgasr}(a) shows the observed $f_{\rm gas}(r)$ profiles
for the six lowest redshift clusters in the sample, for the reference
$\Lambda$CDM cosmology. Although some dispersion in the profiles is
present, particularly at small radii, the profiles tend towards a
common value at $r_{2500}$. Fitting the \fgas~measurements at
$r_{2500}$ for the six lowest-redshift systems with a constant value
we obtain $\fgas=0.113\pm0.003$, with $\chi^2=4.3$ for 5 degrees of
freedom. Fitting the results for all 42 clusters gives
$\fgas=0.1104\pm0.0016$, with $\chi^2=43.5$ for 41 degrees of freedom.

Fig. \ref{fig:fgasr}(b) shows the $f_{\rm gas}(r/r_{2500})$ profiles
for all 42 clusters in the sample. Fitting the data in the range
$0.7-1.2$r$_{2500}$ with a power-law model, we measure
\fgas$=0.1105\pm0.0005(r/r_{2500})^{0.214\pm0.022}$.  Note that the
error bars on the mean $f_{\rm gas}$ measurements quoted above reflect
only the statistical uncertainties in these values. A systematic
uncertainty of $\sim 10-15$ per cent in the global, absolute $f_{\rm gas}$
normalization is also present due to uncertainties in \eg instrument
calibration, X-ray modelling and non-thermal pressure support; this 
must be accounted for in the determination of cosmological constraints
(Section 4.2).

Table \ref{table:fgas} summarizes the results on the X-ray gas mass
fraction for each cluster measured at $r_{2500}$, together with the
$r_{2500}$ values, for the reference $\Lambda$CDM cosmology.
Fig. \ref{fig:fgasz}~shows a comparison of the $f_{\rm gas}$ results,
plotted as a function of redshift, for the reference $\Lambda$CDM
cosmology and a flat, standard cold dark matter (SCDM) cosmology with
$\Omega_{\rm m}=1.0$, $h=0.5$. Whereas the results for the
$\Lambda$CDM cosmology appear consistent with the expectation of a
constant $f_{\rm gas}(z)$ value from non-radiative simulations (\eg
Eke \etal 1998; Crain \etal 2007), as evidenced by the acceptable
$\chi^2$ value quoted above, the results for the reference SCDM
cosmology indicate a clear, apparent drop in $f_{\rm gas}$ as the
redshift increases. The $\chi^2$ value obtained from a fit to the SCDM
data with a constant model, $\chi^2=144$ for 41 degrees of freedom,
shows that the SCDM cosmology is clearly inconsistent with a
prediction that $f_{\rm gas}(z)$ should be constant.

\begin{figure}
\vspace{0.2cm}
\hbox{
\hspace{0.2cm}\psfig{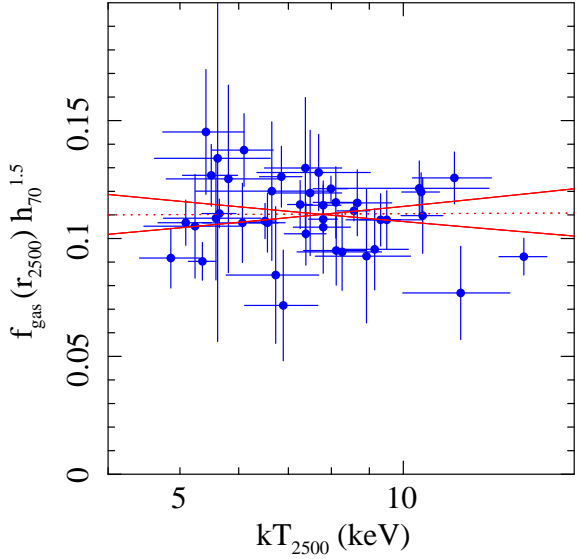}
}
\caption{The X-ray gas mass fraction as a function of mass-weighted
temperature measured within $r_{2500}$ for the reference $\Lambda$CDM
cosmology.  The dotted line shows the best-fitting power law model
which provides a good description of the data ($\chi^2=43.5$ for 40
degrees of freedom) and is consistent with a constant value (slope
$\alpha=0.005\pm0.058$). The solid lines show the $2\sigma$
limits on the slopes allowed by the data.  The figure demonstrates
that \fgas~is essentially independent of temperature for the massive,
dynamically relaxed clusters in the present sample.}\label{fig:fgaskt}
\end{figure}

Table \ref{table:fgas} also lists the mass-weighted temperatures
measured within $r_{2500}$ for each cluster.  Fig. \ref{fig:fgaskt}
shows \fgas~as a function of $kT_{2500}$ for the reference
$\Lambda$CDM cosmology. The dotted line in the figure shows the
best-fitting power law model, \fgas($r_{2500})\propto kT_{2500}^\alpha$,
which provides a good description of the data ($\chi^2=43.5$ for 40
degrees of freedom) and is consistent with a constant value
($\alpha=0.005\pm0.058$).  The solid lines show the $2\sigma$
limits on the steepest and shallowest allowed power law models. It is
clear from the figure that \fgas~is independent of temperature for the
clusters in the present sample.

\subsection{Comparison with previous \fgas~results}
\label{section:compare}

Approximately 0.75\,Ms of the $\sim 1.6$\,Ms of Chandra data used here
were also included in the Allen \etal (2004) study. The current work
includes a re-analysis of those data using improved calibration
information, where available. The \fgas~results from the two studies
show excellent overall agreement: the new \fgas~values are, on
average, $\sim 6$ per cent lower than those reported by Allen \etal
(2004), a difference consistent with expectations given the
modification to the effective area calibration described in
Section~\ref{section:spectra}.

LaRoque \etal (2006) present \fgas~measurements for 38 X-ray luminous
clusters, including 10 of the large, dynamically relaxed systems
studied here. Their best-fit results at $r_{2500}$ are in good overall
agreement with the present work, with their \fgas~values being, on
average, $\sim 6$ per cent higher than those reported here, for the
systems in common.

Pointecouteau \etal (2004) present an analysis of XMM-Newton data for
Abell 478, for which they measure an \fgas~value at $r_{2500}$ of
$0.13\pm0.02$, in good agreement with this work. These authors also
report a value of $0.11$ for Abell 1413, based on the data of Pratt \&
Arnaud (2002), which is consistent with the results reported here.

Vikhlinin \etal (2006) present \fgas~measurements for 13 clusters of
which six are in common with this study. On average, the Vikhlinin
\etal (2006) \fgas~results are $\sim 10$ per cent lower than those
reported here after correcting their values to the same reference
$\Lambda$CDM cosmology.

We note that the statistical uncertainties on the \fgas~measurements
listed in Table~\ref{table:fgas} are, typically, larger than those
reported by other authors. Two contributing factors to this
difference are: 1) that the present analysis does not impose strong
priors on the shapes of the temperature and density profiles in the
clusters through the use of parametric models (the use of such
parameterizations can lead to spuriously tight constraints in cases
where they do not provide an adequate description of the data); and 
2) the \fgas~measurement errors reported here are marginalized over the
uncertainties in all other parameters, including the uncertainties in
$r_{2500}$.

\section{Cosmological analysis}

\subsection{Markov Chain Monte Carlo method}

Our determination of cosmological parameters uses a Markov Chain Monte
Carlo method. We employ a modified version of the CosmoMC
code\footnote{{http://cosmologist.info/cosmomc/}} of Lewis \& Bridle
(2002; see Rapetti \etal 2005, 2007 for details of the enhancements),
which uses a Metropolis-Hastings Markov Chain Monte Carlo (MCMC)
algorithm to explore parameter space.  We run the code on four to
sixteen processors simultaneously, creating multiple chains and
using the Message Passing Interface (MPI) to dynamically update the
proposal matrix based on the covariance of post-burn-in samples. This
leads to a much faster convergence than would be obtained from a
single chain run on a single compute node.

Convergence is assessed using the Gelman-Rubin criterion (Gelman \&
Rubin 1992).  Convergence is deemed acceptable when the ratio of
between-chain to mean-chain variances, $R$, satisfies $R-1<0.1$.  (We
have also visually compared individual chains to ensure that
consistent final results were obtained.) In general, our combined
chains typically have lengths of at least $10^5$ samples and have
$R-1<<0.1$. (For the evolving-$w$ models, $R-1\sim0.1$.)  Conservative
burn-in periods of at least $10000$ samples were allowed for each
chain.

\subsection{Analysis of the \fgas~data: modelling and 
systematic allowances}
\label{section:modelling}

The differences between the shapes of the $f_{\rm gas}(z)$ curves in
Figs. \ref{fig:fgasz}(a) and (b) reflect the dependence of the
measured $f_{\rm gas}$ values on the assumed angular diameter
distances to the clusters.  Under the assumption (Section 1) that
$f_{\rm gas}$ should, in reality, be approximately constant with
redshift, as suggested by non-radiative simulations of large clusters
(Eke \etal 1998; Crain \etal 2007; uncertainties in the predictions
from simulations are discussed below) inspection of
Fig. \ref{fig:fgasz} would clearly favour the $\Lambda$CDM over the
SCDM cosmology.

To determine constraints on cosmological parameters, it is not
necessary to generate $f_{\rm gas}(z)$ data sets for every cosmology
of interest and compare them to the expected behaviour. Rather, one
can fit a single, reference $f_{\rm gas}(z)$ data set with a model that
accounts for the expected apparent variation in $f_{\rm gas}(z)$ as
the underlying cosmology is varied. We choose to work with the
$\Lambda$CDM reference cosmology, although similar results can in
principle be derived for other reference cosmologies.

The model fitted to the reference $\Lambda$CDM data is

\vspace{0.1cm}
\begin{equation}
f_{\rm gas}^{\rm \Lambda CDM}(z) = \frac{ K A \gamma b(z)} {1+s(z) } 
\left( \frac{\Omega_{\rm b}}{\Omega_{\rm m}} \right)
\left[ \frac{d_{\rm A}^{\rm \Lambda CDM}(z)}{d_{\rm A}(z)} \right]^{1.5},
\label{eq:fgas}
\end{equation}
\vspace{0.1cm}

\noindent where $d_{\rm A}(z)$ and 
$d_{\rm A}^{\rm \Lambda CDM}(z)$ are the angular diameter distances to the
clusters in the current test model and reference cosmologies,

\vspace{0.1cm}
\begin{equation}
   d_{A} = \frac{c}{H_0 (1+z)\sqrt{\OK}} \, \sinh \left(\sqrt{\OK}\int_0^z
        {dz\over E(z)}\right),
 \label{eq:DA}
\end{equation}
\vspace{0.1cm}

\noindent with $E(z)$ defined as in Section~\ref{section:demod}.  The
factor $A$ in Equation~\ref{eq:fgas} accounts for the change in angle
subtended by $r_{2500}$ as the underlying cosmology is
varied\footnote{To see the origin of the correction factor $A$, recall
that equation~\ref{eq:fgas} predicts the \fgas~value at the
measurement radius in the reference $\Lambda$CDM cosmology.  This
measurement radius corresponds to a fixed angle $\theta_{2500}^{\rm
\Lambda CDM}$ for each cluster, which will differ slightly from
$\theta_{2500}$, the angle corresponding to $r_{2500}$ for that
cluster in the current test cosmology. The mass contained within
radius $r_{2500}$, $M_{2500}=10^4 \pi r_{2500}^3 \rho_{\rm
crit}/3$. Given that the temperature, and temperature and density
gradients, in the region of $\theta_{2500}$ are likely to be
approximately constant, the hydrostatic equation gives $M_{2500}
\approxpropto r_{2500}$.  Thus, since $\rho_{\rm crit}=3 H(z)^2/8 \pi
G$, we have $r_{2500} \approxpropto H(z)^{-1}$, and the angle spanned
by $r_{2500}$ at redshift $z$, $\theta_{2500}=r_{2500}/d_{\rm A}
\approxpropto (H(z) d_{\rm A})^{-1}$. Since the \fgas~profiles follow
a smooth power law form in the region of $\theta_{2500}$, the ratio of
the model \fgas~value at $\theta_{2500}^{\rm \Lambda CDM}$ to that at
$\theta_{2500}$ can be described by
Equation~\ref{eqn:angcorrection}.}:

\vspace{0.1cm}
\begin{equation}
A= \left( \frac{ \theta_{2500}^{\rm \Lambda CDM}}{\theta_{2500}} \right)^\eta \approx
\left( \frac{ H(z) d_{\rm A}(z)~~~~~~~} { \left[ H(z) d_{\rm A}(z)\right] ^{\rm \Lambda CDM}} \right)^\eta. 
\label{eqn:angcorrection}
\end{equation}
\vspace{0.1cm}

\noindent Here, $\eta$ is the slope of the \fgas($r/r_{2500}$) data in
the region of $r_{2500}$, as measured for the reference $\Lambda$CDM
cosmology. For simplicity, we use the best-fit average slope of
$\eta=0.214\pm0.022$ determined from a fit to the whole sample over
the range $0.7<r/r_{2500}<1.2$ (Section~\ref{section:fgasres}) and
marginalize over the slope uncertainty. This angular correction
factor, which is close to unity for all cosmologies and redshifts
of interest, has not been employed in previous studies and, indeed,
can be neglected without significant loss of accuracy for most work.
Nevertheless, we include it here for completeness and note that its
inclusion leads to slightly tighter constraints on dark energy than
would otherwise be obtained.

The parameter $\gamma$ in equation~\ref{eq:fgas} models non-thermal
pressure support in the clusters. Based on hydrodynamical simulations,
Nagai \etal (2007a) estimate a bias of $\sim 9$ per cent in
\fgas~measurements at $r_{2500}$ for relaxed clusters. This bias
originates primarily from subsonic motions in the intracluster gas
and, as discussed by those authors (see also
Section~\ref{section:scatter}), can be regarded as an upper limit,
given observational indications that the gas viscosity in real
clusters appears likely to exceed that modelled in the simulations.
For the large, relaxed clusters and measurement radii of interest
here, non-thermal pressure support due to cosmic rays (Pfrommer \etal
2007) and magnetic fields (Dolag \& Schindler 2000) is expected to be
small. Based on these considerations, our default analysis assumes 
a uniform prior of $1.0<\gamma<1.1$, although we also consider the
case where the non-thermal pressure support may be up to twice
as large \ie  $1.0<\gamma<1.2$.

The parameter $s(z)=s_0(1+\alpha_{\rm s}z)$ in equation~\ref{eq:fgas}
models the baryonic mass fraction in stars.  As discussed in
Section~\ref{section:opt}, we include a 30 per cent Gaussian
uncertainty on $s_0$, such that $s_0=(0.16\pm0.05)h_{70}^{0.5}$, and a
20 per cent uniform prior on $\alpha_{\rm s}$, such that
$-0.2<\alpha_{\rm s}<0.2$, allowing for evolution in the stellar
baryonic mass fraction of $\pm 20$ per cent per unit redshift
interval.

\begin{figure}
\vspace{0.5cm} \hbox{
\hspace{-0.1cm}\psfig{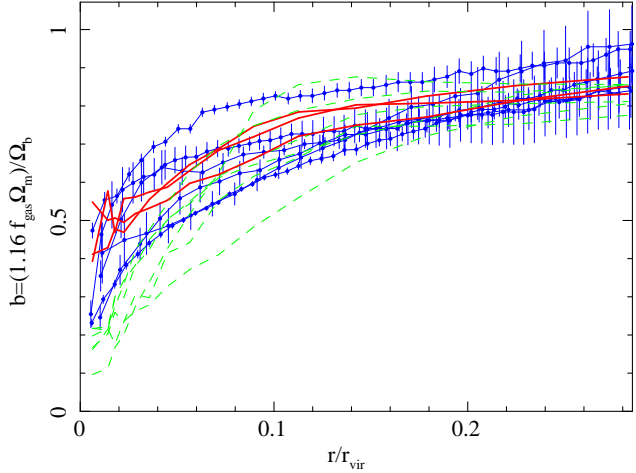}}
\caption{The X-ray depletion or bias factor, $b$ (\ie the enclosed
baryon fraction relative to the universal value) as a function of
radius, in units of the virial radius $r_{\rm vir}$, from the
simulations of Eke \etal (1998).  The simulated clusters have similar
masses to the systems studied here. The results (at zero redshift) for
the three most dynamically relaxed clusters in the simulations are
shown as bold red curves. Less relaxed simulated clusters are shown
as dashed green curves.  The Chandra observations for the six lowest
redshift clusters in the \fgas~sample are plotted as blue circles, with
error bars.  (The Chandra profiles are identical to those shown in
Fig~\ref{fig:fgasr}, but are scaled assuming $\Omega_{\rm m}=0.27$,
$\Omega_{\rm b}=0.0413$ and $r_{2500}=0.25\,r_{\rm vir}$.)  The
agreement between the observed and predicted profiles argues that the
non-radiative simulations provide a reasonable approximation for
the purpose of predicting the baryonic mass distributions.}\label{fig:eke}
\end{figure}

The factor $b(z)=b_0(1+\alpha_{\rm b}z)$ is the `depletion' or `bias'
factor \ie the ratio by which the baryon fraction measured at
$r_{2500}$ is depleted with respect to the universal mean; such
depletion is a natural consequence of the thermodynamic history of the
gas. The non-radiative simulations of hot, massive clusters published
by Eke \etal (1998; see also Crain \etal 2007) give $b_0=0.83\pm0.04$
at $r_{2500}$, and are consistent with no redshift evolution in $b$
for $z<1$.  We use these simulations as a benchmark because other
simulations that include cooling currently tend to significantly
over-produce young stars in the largest galaxies (see \eg Balogh \etal
2001), which is problematic for the prediction of $b(z)$.  We note
also the good agreement between the observed, scaled $f_{\rm gas}(r)$
profiles determined from the Chandra data and the $b(r)$ profiles for
the three most relaxed clusters in the simulations of Eke \etal (1998;
the red curves in Fig~\ref{fig:eke}); this suggests that the
non-radiative simulations provide a useful approximation for the
purpose of predicting $b(z)$. (The profiles for the less relaxed
simulated clusters are shown as dashed green curves in the figure.)
Nevertheless, to account for systematic uncertainties in the
predictions of $b(z)$, we include a conservative 20 per cent uniform
prior on $b_0$, such that $0.65<b_0<1.0$, and allow for moderate,
systematic evolution in $b(z)$ over the observed redshift range,
setting $-0.1<\alpha_b<0.1$. This encompasses a range of evolution
allowed by recent simulations including various approximations to the
detailed baryonic physics (\eg Kay \etal 2004, Ettori \etal 2006,
Crain \etal 2007, Nagai \etal 2007a).

The factor $K$ in equation~\ref{eq:fgas} is a `calibration' constant
that parameterizes residual uncertainty in the accuracy of the
instrument calibration and X-ray modelling. Contributing factors
include uncertainty in the instrument effective area, variations
in element abundance ratios, modelling the
effects of gas clumping and asphericity (the latter effects are
expected to be small for large, relaxed clusters; Nagai \etal
2007a. See also Piffaretti, Jetzer \& Schindler 2003, Gavazzi 2005).
We conservatively include a 10 per cent Gaussian uncertainty in $K$ to
model the combined effect of these factors, such that $K=1.0\pm0.1$. The
small intrinsic dispersion in \fgas~values
(Section~\ref{section:scatter}) means that Malmquist bias is
expected to have a negligible effect on the derived cosmological
parameters. Uncertainties associated with other systematic factors 
are expected to be negligible in comparison to the allowances 
listed above.
 
In cases where the Chandra \fgas~data are not combined with CMB data,
we include simple Gaussian priors on $\Omega_{\rm b}h^2$ and $h$.  Two
separate sets of priors were used: `standard' priors with $\Omega_{\rm
b}h^2=0.0214\pm0.0020$ (Kirkman \etal 2003) and $h=0.72\pm0.08$
(Freedman \etal 2001), and `weak' priors in which the nominal
uncertainties were tripled to give $\Omega_{\rm b}h^2=0.0214\pm0.0060$
and $h=0.72\pm0.24$.  In cases where the CMB data are included, no
priors on $\Omega_{\rm b}h^2$ or $h$ are needed or used.
The complete set of standard priors and allowances included in 
the $f_{\rm gas}$ analysis are summarized in Table~\ref{table:sys}. 

Finally, we note how inspection of equation~\ref{eq:fgas} can provide
useful insight into the strength of the \fgas~experiment.  The
pre-factors before the square brackets shows how the $normalization$
of the $f_{\rm gas}(z)$ curve is used to constrain $\Omega_{\rm m}$,
given prior information on $\Omega_{\rm b}$, $h$, $K$, $\gamma$, $b$
and $s$.  The ratio of distances inside the square brackets (and to a
small extent the angular correction factor) shows how the $shape$ of
the $f_{\rm gas}(z)$ curve constrains the geometry of the Universe and
therefore dark energy.  The combination of information from both the
normalization and shape {\it breaks the degeneracy} between
$\Omega_{\rm m}$ and the dark energy parameters in the distance
equations.

\subsection{Other data used in the analysis}
\label{section:otherdata}

In addition to the analysis of the Chandra \fgas~data alone, we have
examined the improved constraints on cosmological parameters that can
be obtained through combination of the \fgas~data with CMB and SNIa
studies.

Our analysis of CMB observations uses the three-year WMAP temperature
(TT) data for multipoles $l<1000$ (Hinshaw \etal 2007; Spergel \etal
2007) and temperature-polarization (TE) data for $l<450$ (Page \etal
2007). We use the October 2006 version of the WMAP likelihood code
available from {\it
http://lambda.gsfc.nasa.gov/product/map/current/m\_sw.cfm}.  Like most
authors, we have ignored the small contribution to the TT data
expected to arise from the Sunyaev-Z'eldovich (SZ) effect in clusters
and groups (\eg Komatsu \& Seljak 2002) and do not account for
gravitational lensing of the CMB (Lewis \& Challinor 2006), which has
a negligible effect on the derived cosmological parameters. To extend
the analysis to higher multipoles (smaller scales), we also include
data from the Cosmic Background Imager (CBI; Mason \etal 2003; Pearson
\etal 2003), the Arcminute Cosmology Bolometer Array Receiver (ACBAR;
Kuo \etal 2003) and BOOMERanG (Jones \etal 2006; Montroy \etal 2006;
Piacentini \etal 2005), as incorporated into the current version of the
CosmoMC code (Lewis \& Bridle 2002). We use a modified version of CAMB
(Lewis, Challinor \& Lasenby 2000) to calculate CMB power spectra,
which includes a consistent treatment of the effects of dark energy
perturbations for evolving$-w$ models (Rapetti \etal 2005; we assume 
that the sound speed in the dark energy fluid is equal to the speed of light).

Our analysis of SNIa data uses two separate supernova samples. In the
first case, we use the compilation of Davis \etal (2007) which includes
results from the ESSENCE survey (60 targets; Wood-Vasey \etal 2007, 
Miknaitis \etal 2007),
the SNLS first year data (57 targets; Astier \etal 2006), 45 nearby
supernovae (Jha \etal 2007) and the 30 high-redshift supernovae
discovered by HST and reported by Riess \etal (2007) for which a
`gold' rating was awarded. This sample includes 192 SNIa in total.
The second supernova sample is the full `gold' sample of Riess \etal
(2007) which totals 182 SNIa, including the HST-discovered objects.
For both samples we marginalize analytically over the absolute
normalization of the distance moduli.

\subsection{Dark Energy models}
\label{section:demod}

We have considered three separate dark energy models in the analysis:
1) standard $\Lambda$CDM, for which the dark energy equation of state
$w=-1$; 2) a model that allows any constant dark energy
equation of state, including `phantom' models with $w<-1$; 3) a
model in which the dark energy equation of state is allowed to evolve
as

\vspace{0.1cm}
\begin{equation}
w=\frac{w_{\rm et}z+w_{\rm 0}z_{\rm t}}{z+z_{\rm t}}=\frac{w_{\rm et}(1-a)a_{\rm t}+w_{\rm
0}(1-a_{\rm t})a} {a(1-2a_{\rm t})+a_{\rm t}}, 
\label{eq:evolve}
\end{equation}
\vspace{0.1cm}

\noindent where $a=1/(1+z)$ is the scale factor, $w_0$ and $w_{\rm
et}$ are the equation of state at late (present day) and early times,
and $z_{\rm t}$ and $a_{\rm t}$ are the redshift and scale factor at
the transition between the two, respectively (Rapetti \etal 2005; see
also Chevallier \& Polarski 2001; Linder 2003; Corasaniti \etal 2003; Linder
2007).  We employ a uniform prior on the transition scale factor such
that $0.5<a_{\rm t}<0.95$. As discussed by Rapetti \etal (2005), this
model is both more general and more applicable to current data, which
primarily constrain the properties of dark energy at redshifts $z<1$,
than models which impose a transition redshift $z=1$, \eg
$w(a)=w_0+w_a(1-a)$.

Energy conservation of the dark energy fluid leads to an evolution
of the energy density with scale factor 

\vspace{0.1cm}
\begin{equation}
\rho_{\rm de}(a)=\rho_{{\rm de,}0}a^{-3}e^{-3\int_{1}^{a}{\frac{w(a')}{a'}da'}},\; 
\label{eqn:conservation}
\end{equation}
\vspace{0.1cm}

\noindent where $\rho_{{\rm de,}0}$ is the energy density of the dark 
energy fluid today. Using the parameterization of equation
(\ref{eq:evolve}) we obtain

\vspace{0.1cm}
\begin{equation}
\int_{1}^{a}{\frac{w(a')}{a'}da'}=w_{\rm et}\ln a + 
(w_{\rm et}-w_{\rm 0})g(a;a_{\rm t})\; ,
\label{eqn:intwa}
\end{equation}
\vspace{0.1cm}

\noindent with

\begin{equation}
g(a;a_{\rm t})=\left(\frac{1-a_{\rm t}}{1-2a_{\rm t}}\right)\ln \left( \frac{ 1-a_{\rm t} }{a(1-2a_{\rm t})+a_{\rm t}} \right)\;.
\label{eqn:intwa2}
\end{equation}
\vspace{0.1cm}

\noindent The Friedmann equation, which relates the first time
derivative of the scale factor of the Universe to the total density,
can be conveniently expressed as $({\dot
a}/a)^2\,$=$\,H(a)^2\,$=\,$H_0^2E(a)^2$, with

\vspace{0.1cm}
\begin{equation}
E(a) = \sqrt{ \OM a^{-3} + \ODE f(a) + \OK a^{-2}}.
\label{eq:GR4}
\end{equation}
\vspace{0.1cm}

\noindent Here $\OK$ is the curvature, $\ODE$ is the dark energy
density and $f(a)$ is its redshift dependence. (Note that we have
ignored the density contributions from radiation and relativistic
matter in this expression, although they are included in the
analysis.)  For our most general dark energy parameterization
(Equation \ref{eq:evolve})

\vspace{0.1cm}
\begin{equation}
f(a) = a^{-3(1+w_{\rm et})} e^{-3(w_{\rm et}-w_{\rm 0})g(a;a_{\rm t})}.
\end{equation}
\vspace{0.1cm}

\noindent For $\Lambda$CDM cosmologies, the dark energy density is
constant and $f(a)=1$.  For $w < -1$ the dark energy density 
increases with time.  For constant $w$ models with $w < -1/3$, dark
energy accelerates the expansion of the universe. (The results from a
purely kinematic modelling of the data, which does not rely on the
Friedmann equation and is independent of the assumptions of General
Relativity, are discussed by Rapetti \etal 2007).

Our combined analysis of Chandra \fgas, SNIa and CMB data therefore
has up to ten interesting parameters: the physical dark matter and
baryon densities in units of the critical density, the curvature
$\OK$, the ratio of the sound horizon to the angular diameter distance
for the CMB (Kosowsky, Milosavljevic \& Jimenez 2002), the amplitude
of the scalar power spectrum, the scalar spectral index, the optical
depth to reionization, and up to three parameters associated with the
dark energy equation of state: $w_0$, $w_{\rm et}$ and $a_{\rm t}$.
In all cases, we assume an absence of both tensor components and
massive neutrinos and, for the analysis of the CMB data alone, include
a wide uniform prior on the Hubble parameter, $0.2 < h <2.0$. (Tests
in which tensor components are included with $\Lambda$CDM models lead
to similar results on dark energy, but take much longer to compute.)

\section{Constraints on cosmological parameters}

\begin{table}
\begin{center}
\caption{Summary of the standard systematic allowances and priors included 
in the Chandra \fgas~analysis. The priors on $\OBH$ and $h$ 
(Kirkman \etal 2003, Freedman \etal 2001) are used when the CMB data 
are not included. We have also 
examined the case where the allowance for 
non-thermal pressure support has been doubled 
\ie  $1.0<\gamma<1.2$ (see text for details).
}\label{table:sys}
\begin{tabular}{ c c c }
                             & Parameter        & Allowance \\
\hline
Calibration/Modelling        & $K$              & $1.0\pm0.1$ (Gaussian) \\
Non-thermal pressure         & $\gamma$         & $1.0<\gamma<1.1$ \\
Gas depletion: norm.         & $b_0$            & $0.65<b_0<1.0$ \\
Gas depletion: evol.         & $\alpha_{\rm b}$ & $-0.1<\alpha_{\rm b}<0.1$ \\
Stellar mass: norm.          & $s_0$            & $0.16\pm0.048$ (Gaussian) \\
Stellar mass: evol.          & $\alpha_{\rm s}$ & $-0.2<\alpha_{\rm s}<0.2$ \\
$f_{\rm gas}(r\sim r_{2500})$ slope & $\eta$    & $0.214\pm0.022$ (Gaussian)  \\
&& \\             
Standard prior $\OBH$        & $\OBH$           & $0.0214\pm0.0020$ \\
Standard prior $h$           & $h$              & $0.72\pm0.08$ \\ 
Weak prior $\OBH$            & $\OBH$           & $0.0214\pm0.0060$ \\
Weak prior $h$               & $h$              & $0.72\pm0.24$ \\ 
\hline                      
\end{tabular}
\end{center}
\end{table}

\subsection{Constraints on $\Omega_{\rm m}$ from the low$-z$ \fgas~data}
\label{section:lowz}

\begin{figure}
\vspace{0.5cm}
\hbox{
\hspace{0.2cm}\psfig{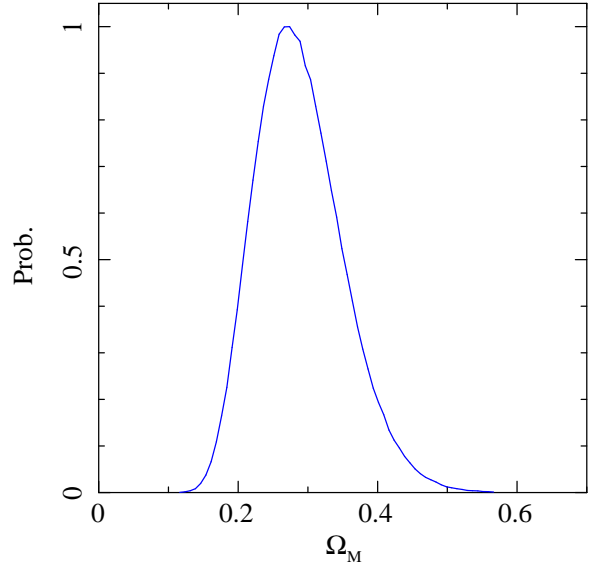}
} 
\caption{The marginalized constraints on $\OM$ from the Chandra
$f_{\rm gas}$ data for the six lowest redshift clusters, using the
non-flat $\Lambda$CDM model and standard priors on $\OBH$ and $h$.
Uncertainties due to the evolution in $b$ and $s$ and the nature of
the dark energy component are negligible in the analysis (although 
allowances for these uncertainties are included). We obtain a
marginalized result $\OM=0.28\pm0.06$ (68 per cent confidence
limits).}
\label{fig:om6low}
\end{figure}

In the first case, we have used the Chandra \fgas~data for only the
six, lowest redshift clusters in the sample, with $z\approxlt 0.15$,
to constrain the mean matter density of the Universe.  The restriction
to low-$z$ clusters minimizes correlated uncertainties associated with
the nature of the dark energy component (dark energy has only a very
small effect on the space-time metric over this redshift range; we
employ a broad uniform prior such that $0.0<\OL<2.0$) and renders
negligible uncertainties associated with the evolution of the
depletion factor and stellar baryonic mass fraction ($\alpha_{\rm b}$
and $\alpha_{\rm s}$).  Fig.~\ref{fig:om6low} shows the marginalized
constraints on $\OM$ for a $\Lambda$CDM model with free curvature,
using the standard priors on $\OBH$ and $h$, for which we obtain a
result of $\OM=0.28\pm0.06$. The full set of conservative systematic
allowances, as described in Table~\ref{table:sys}, were included.

The result on $\OM$ from the six lowest redshift clusters is in good
agreement with that obtained for the whole sample, as discussed below.
It is also consistent with the result on $\OM$ found from an analysis
of all clusters $except$ the six-lowest redshift systems,
$\OM=0.29\pm0.06$, \ie the six-lowest redshift clusters do not
dominate the $\OM$ constraints. Note that the error bars on $\OM$ are
dominated by the widths of the priors on $\OBH$ and $h$ and the
magnitudes of the systematic allowances on $K$, $b$ and $\gamma$,
which are all at the $\sim 10-20$ per cent level. In contrast, the
statistical uncertainty in the normalization of the \fgas(z)~curve is
small (Section~\ref{section:fgasres2}) and has a negligible impact on
the $\OM$ results.

The result on $\OM$ is consistent with previous findings based on
\fgas~data (see references in Section 1) and independent constraints
from the CMB (\eg Spergel \etal 2007), galaxy redshift surveys (\eg
Eisenstein \etal 2005) and other leading cosmological data. Note that
the agreement in cosmological parameters determined from the \fgas~and
CMB data argues against any unusual depletion of baryons within
$r_{2500}$ in hot, relaxed clusters (see \eg the discussions in Ettori
2003, Afshordi \etal 2007 and McCarthy, Bower \& Balogh 2007)

\subsection{Constraints on the $\Lambda$CDM model using
the \fgas~(+CMB+SNIa) data}

\begin{figure}
\vspace{0.5cm}
\hbox{
\hspace{-0.2cm}\psfig{figure=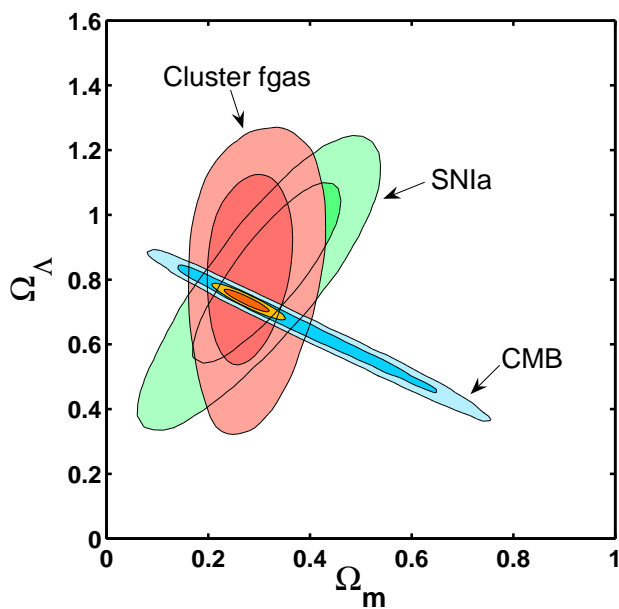,width=.51\textwidth,angle=0}
} 
\caption{The 68.3 and 95.4 per cent (1 and 2 $\sigma$) confidence
constraints in the $\OM,\OL$ plane for the Chandra $f_{\rm gas}$ data
(red contours; standard priors on $\OBH$ and $h$ are used). Also shown
are the independent results obtained from CMB data (blue contours)
using a weak, uniform prior on $h$ ($0.2<h<2$), and SNIa data (green
contours; the results for the Davis \etal 2007 compilation are shown). The
inner, orange contours show the constraint obtained from all three
data sets combined (no external priors on $\OBH$ and $h$ are used).  A
$\Lambda$CDM model is assumed, with the curvature included as a free
parameter.}\label{fig:lcdm_cont}
\end{figure}

We next extended our analysis to measure $\OM$ and $\OL$ for a
non-flat $\Lambda$CDM model using the Chandra \fgas~data for the full
sample of 42 clusters. The results are shown as the red contours in
Fig.~\ref{fig:lcdm_cont}.  Using the systematic allowances summarized
in Table~\ref{table:sys} and the standard priors on $\OBH$ and $h$, we
measure $\OM=0.27\pm0.06$ and $\OL=0.86\pm0.19$, (68 per cent
confidence limits) with $\chi^2=41.5$ for 40 degrees of freedom. The
low $\chi^2$ value obtained is important and indicates that the model
provides an acceptable description of the data (see
Section~\ref{section:scatter} below). The result on $\OM$ is in
excellent agreement with that determined from the six lowest redshift
clusters only (Section~\ref{section:lowz}). The result is also
consistent with the value reported by Allen \etal (2004) using the
previous release of \fgas~data, although the more conservative systematic
allowances included here lead to the quoted uncertainties in $\OM$
being larger by $\sim 50$ per cent.

\begin{figure}
\vspace{0.5cm}
\hbox{
\hspace{0.2cm}\psfig{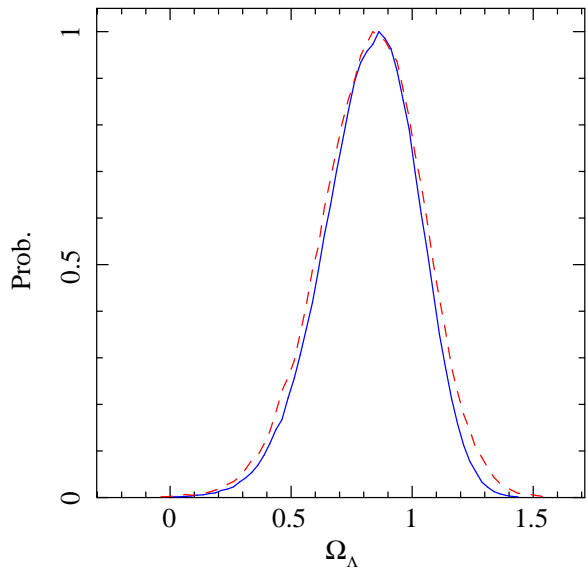}
} 
\caption{The marginalized constraints on $\OL$ determined from the
Chandra $f_{\rm gas}$ data using the non-flat $\Lambda$CDM model and
standard (solid curve) and weak (dashed curve) priors on
$\OBH$ and $h$. The \fgas~data provide a detection of the effects of
dark energy at the $\sim 99.99$ per cent confidence
level.}\label{fig:lcdm_marg}
\end{figure}

Fig.~\ref{fig:lcdm_marg} shows the marginalized constraints on
$\Omega_{\Lambda}$ obtained using both the standard and weak priors on
$\OBH$ and $h$. We see that using only the weak priors
($\OBH=0.0214\pm0.0060$, $h=0.72\pm0.24$), the $f_{\rm gas}$ data
provide a clear detection of the effects of dark energy on the
expansion of the Universe, with $\OL=0.86\pm0.21$: a model with
$\Omega_{\Lambda}\leq0$ is ruled out at $\sim 99.98$ per cent
confidence. (Using the standard priors on $\OBH$ and $h$, a model with
$\Omega_{\Lambda}\leq0$ is ruled out at $99.99$ per cent confidence;
Table~\ref{table:results}). The significance of the detection of dark
energy in the \fgas~data is comparable to that of current SNIa studies
(\eg Riess \etal 2007; Wood-Vasey \etal 2007). The ~\fgas~data provide
strong, independent evidence for cosmic acceleration.

In contrast to the $\OM$ constraints, the error budget for $\OL$
includes significant contributions from both statistical and
systematic sources. From the analysis of the full sample of 42
clusters using the standard priors on $\OBH$ and $h$, we find
$\OL=0.86\pm0.19$; the error bar is comprised of 
approximately $\pm0.15$ statistical
error and $\pm0.12$ systematic uncertainty.  Thus, whereas improved
measurements of $\OM$ from the \fgas~method will require 
additional information leading to tighter priors and systematic allowances,
significant improvements in the precision of the dark energy
constraints should be possible simply by gathering more data (\eg
doubling the present \fgas~data set).

Fig.~\ref{fig:lcdm_cont} also shows the constraints on $\OM$ and $\OL$
obtained from the CMB (blue contours) and SNIa (green contours) data
(Section~\ref{section:otherdata}). The agreement between the results
for the independent data sets is excellent and motivates a combined
analysis.  The inner, orange contours in Fig.~\ref{fig:lcdm_cont} show
the constraints on $\OM$ and $\OL$ obtained from the combined $f_{\rm
gas}$+CMB+SNIa data set. We obtain marginalized 68 per cent confidence
limits of $\Omega_{\rm m}=0.275\pm0.033$ and
$\Omega_{\Lambda}=0.735\pm0.023$. Together, the $f_{\rm gas}$+CMB+SNIa
data also constrain the Universe to be close to geometrically flat:
$\OK=-0.010\pm0.011$.  No external priors on $\OBH$ and $h$ are used
in the analysis of the combined \fgas+CMB+SNIa data (see also
Section~\ref{section:degenbreak}).

Finally, we have examined the effects of doubling the
allowance for non-thermal pressure support in the clusters
\ie setting $1.0<\gamma<1.2$. For the analysis of the \fgas~data
alone, this boosts the best-fit value of $\OM$ by $\sim
5$ per cent but leaves the results on dark energy unchanged. This can
be understood by inspection of equation~\ref{eq:fgas} and recalling
that the constraint on $\OM$ is determined primarily from the
normalization of the $\fgas$ curve, whereas the constraints on dark
energy are driven by its shape (Section~\ref{section:modelling}). For
the combined \fgas+CMB+SNIa data set, doubling the width of the
allowance on $\gamma$ has a negligible impact on the results, since in
this case the value of $\OM$ is tightly constrained by the combination
of data sets.

\begin{table*}
\begin{center}
\caption{Summary of the constraints on cosmological parameters
determined from the Chandra \fgas~data and complementary data sets.
Error bars reflect the combined statistical and systematic
uncertainties, incorporating the allowances and priors described in
Section~\ref{section:modelling}.  For the low-$z$ \fgas~data
($z<0.15$), the constraint on $\OM$ is almost independent of the
details of the dark energy component (Section~\ref{section:lowz}).
The SNIa(1) and SNIa(2) labels denote the supernovae samples of Davis
\etal (2007) and Riess \etal (2007), respectively
(Section~\ref{section:otherdata}).}
\label{table:results}
\vskip 0 truein
\begin{tabular}{ c c c c c c c c }
&&&&&&&  \\
\multicolumn{1}{c}{} &
\multicolumn{1}{c}{} &
\multicolumn{1}{c}{} &
\multicolumn{1}{c}{} &
\multicolumn{4}{c}{COSMOLOGICAL CONSTRAINTS} \\

Data              &      Model                     & $\OBH$, $h$ Priors  & ~ & $\OM$                  &  $\ODE$       & $w_0$   & $w_{\rm et}$     \\
\hline
low-$z$ \fgas~    &  $\Lambda$CDM (0$<\OL<$2.0)      & standard  & ~ & $0.28\pm0.06$          &    ---         &  ---                      &    ---           \\
\fgas             &  $\Lambda$CDM                    & standard  & ~ & $0.27\pm0.06$          & $0.86\pm0.19$  &  ---                      &    ---           \\
\fgas             &  $\Lambda$CDM                    & weak      & ~ & $0.27\pm0.09$          & $0.86\pm0.21$  &  ---                      &    ---           \\
\fgas+CMB         &  $\Lambda$CDM                    & none      & ~ & $0.28\pm0.06$          & $0.73\pm0.04$  &  ---                      &    ---           \\
\fgas+CMB+SNIa(1) &  $\Lambda$CDM                    & none      & ~ & $0.275\pm0.033$        & $0.735\pm0.023$&  ---                      &    ---           \\
&&&&&&& \\             
\fgas             &  constant $w$ (flat)             & standard  & ~ & $0.28\pm0.06$          &    ---         & $-1.14^{+0.27}_{-0.35}$   &    ---           \\
\fgas             &  constant $w$ (flat)             & weak      & ~ & $0.29\pm0.09$          &    ---         & $-1.11^{+0.31}_{-0.45}$   &    ---           \\
\fgas+CMB         &  constant $w$ (flat)             & none      & ~ & $0.243\pm0.033$        &    ---         & $-1.00\pm0.14$            &    ---           \\
\fgas+CMB+SNIa(1) &  constant $w$ (flat)             & none      & ~ & $0.253\pm0.021$        &    ---         & $-0.98\pm0.07$            &    ---           \\
\fgas+CMB+SNIa(1) &  constant $w$                    & none      & ~ & $0.310\pm0.052$        & $0.713\pm0.036$& $-1.08^{+0.13}_{-0.19}$   &    ---           \\
&&&&&&& \\             
\fgas+CMB+SNIa(1) &  evolving $w$ (flat)             & none      & ~ & $0.254\pm0.022$        &    ---         & $-1.05^{+0.31}_{-0.26}$   & $-0.83^{+0.48}_{-0.43}$  \\
\fgas+CMB+SNIa(1) &  evolving $w$                    & none      & ~ & $0.29^{+0.09}_{-0.04}$ & $0.71^{+0.04}_{-0.05}$ & $-1.15^{+0.50}_{-0.38}$   & $-0.80^{+0.70}_{-1.30}$  \\
\fgas+CMB+SNIa(2) &  evolving $w$ (flat)             & none      & ~ & $0.287\pm0.026$        &    ---         & $-1.19^{+0.29}_{-0.35}$   & $-0.33^{+0.18}_{-0.34}$  \\

\hline                      
\end{tabular}
\end{center}
\end{table*}

\subsection{Scatter in the \fgas~data}
\label{section:scatter}

Hydrodynamical simulations suggest that the intrinsic dispersion in
\fgas~measurements for the largest, dynamically relaxed galaxy
clusters should be small. Nagai \etal (2007a) simulate and analyze
mock X-ray observations of galaxy clusters (including cooling and
feedback processes), employing standard assumptions of spherical
symmetry and hydrostatic equilibrium and identifying relaxed systems
based on X-ray morphology in a similar manner to that employed here.
For relaxed clusters, these authors find that \fgas~measurements at
$r_{2500}$ are biased low by $\sim 9$ per cent, with the bias
primarily due to non-thermal pressure support provided by subsonic
bulk motions in the intracluster gas. They measure an intrinsic dispersion in the
\fgas~measurements of $\sim 6$ per cent, with an indication that the
scatter may be even smaller for analyses limited to the hottest,
relaxed systems with $kT\approxgt 5$keV. Nagai \etal (2007a) also
suggest that the true bias and scatter may be yet smaller if their simulations
have underestimated the viscosity of the X-ray emitting
gas.\footnote{Recent work on the morphologies of X-ray cavities and
H$\alpha$ filaments suggest a relatively high gas viscosity (low
Reynolds number) in nearby cluster cores (Fabian \etal 2003a,b,2005,
Ruszkowski, Br\"uggen \& Begelman 2004, Reynolds \etal 2005).} In
contrast, for $unrelaxed$ simulated clusters, Nagai \etal (2007a) find
that \fgas~ measurements are biased low by on average 27 per cent
with an intrinsic dispersion of more than 50 per cent. Thus, the
dispersion in \fgas~measurements for unrelaxed clusters is expected to
be an order of magnitude larger than for relaxed systems. This is in
agreement with the measurement of very low intrinsic systematic
scatter in the \fgas~data for relaxed clusters reported here (see
below) and the much larger scatter measured in previous works that
included no such restriction to relaxed clusters.  Earlier,
non-radiative simulations by Eke \etal (1998) also argued for a small
intrinsic scatter in \fgas, at the few per cent level, for large,
relaxed clusters (see also Crain \etal 2007). Likewise, Kay \etal
(2004) measure a small intrinsic dispersion in \fgas~measurements from
simulations including cooling and moderate star formation.

The expectation of a small intrinsic dispersion in the
\fgas~measurements for hot, dynamically relaxed clusters is strikingly
confirmed by the present data.  Even without including the allowances
for systematic uncertainties associated with $\gamma$, $b_0$,
$\alpha_{\rm b}$, $s$ and $\alpha_{\rm s}$ described in
Table~\ref{table:sys} (\ie keeping only the 10 per cent systematic
uncertainty on the overall normalization, as described by $K$) the
best-fitting non-flat $\Lambda CDM$ model gives an acceptable
$\chi^2=41.9$ for 40 degrees of freedom, when fitting the full
\fgas~sample. (The $\chi^2$ drops only to 41.5 with the full set of
systematic allowances included; this small change in $\chi^2$
illustrates the degeneracies between the systematic allowances and
model parameters.)  The acceptable $\chi^2$ for the best-fitting model
rules out the presence of significant intrinsic, systematic scatter in
the current \fgas~data. This absence of systematic scatter is observed
despite the fact that the root-mean-square scatter in the \fgas~data
is only 15 per cent. Moreover, the rms scatter is dominated by those
measurements with large statistical uncertainties; the weighted mean
scatter of the \fgas~data about the best-fit $\Lambda$CDM model is
only 7.2 per cent, which corresponds to only $7.2/1.5=4.8$ per cent in
distance.

\subsection{Constraints on the constant $w$ model using
the \fgas~(+CMB+SNIa) data}

\begin{figure}
\vspace{0.5cm}
\hbox{
\hspace{-0.2cm}\psfig{figure=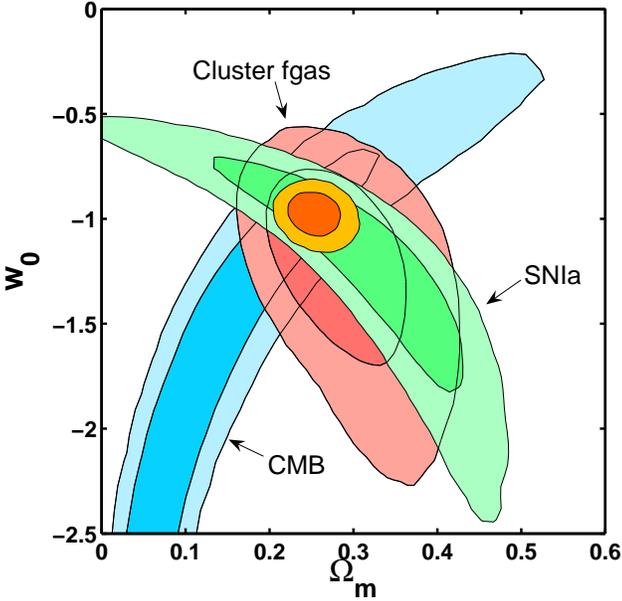,width=.51\textwidth,angle=0}
} 
\caption{The 68.3 and 95.4 per cent (1 and $2\sigma$) confidence
constraints in the $\OM,w$ plane obtained from the analysis of the
Chandra $f_{\rm gas}$ data (red contours) using standard priors on
$\OBH$ and $h$. Also shown are the independent results obtained from
CMB data (blue contours) using a weak, uniform prior on $h$
($0.2<h<2.0$) and SNIa data (green contours; Davis \etal 2007). The inner, 
orange contours show the constraint obtained from all three data sets
combined: $\OM=0.253\pm0.021$ and $w=-0.98\pm0.07$ (68 per cent
confidence limits). No external priors on $\OBH$ and $h$ are used
when the data sets are combined. A flat cosmology with a constant dark energy
equation of state parameter $w$ is assumed.}\label{fig:w_const}
\end{figure}

We have next examined the ability of our data to constrain the dark
energy equation of state parameter, $w$. In the first case, we
examined a geometrically flat model in which $w$ is constant with
time. Fig.~\ref{fig:w_const} shows the constraints in the $\OM$, $w$
plane for this model using the Chandra \fgas~data and standard
priors/allowances (red contours), the CMB data (blue contours) and
SNIa data (green contours).  The different parameter degeneracies in
the data sets are clearly evident. For the \fgas~data alone, we
measure $\OM=0.28\pm0.06$ and $w=-1.14^{+0.27}_{-0.35}$.

The results for the three data sets shown in Fig.~\ref{fig:w_const}
are each, individually, consistent with the $\Lambda$CDM model
($w=-1$). The consistent nature of these constraints again motivates a
combined analysis of the data, shown as the small, central (orange)
contours. For the three data sets combined, we measure
$\OM=0.253\pm0.021$ and $w=-0.98\pm0.07$ (68 per cent confidence
limits). No priors on $\OBH$ and $h$ are required or used in the
combined \fgas+CMB+SNIa analysis.  The constraints on $w$ from the
combined data set are significantly tighter than 10 per cent.

We note that our analysis accounts for the effects of dark energy
perturbations, which must exist for dark energy models other than
$\Lambda$CDM; neglecting the effects of such perturbations can lead to
spuriously tight constraints (See Rapetti \etal 2005 for details).

\subsection{Constraints on the evolution of $w$ from the 
combined \fgas+CMB+SNIa data}
\label{section:evol}

\begin{figure*}
\vspace{0.2cm}
\hbox{
\hspace{0.2cm}\psfig{figure=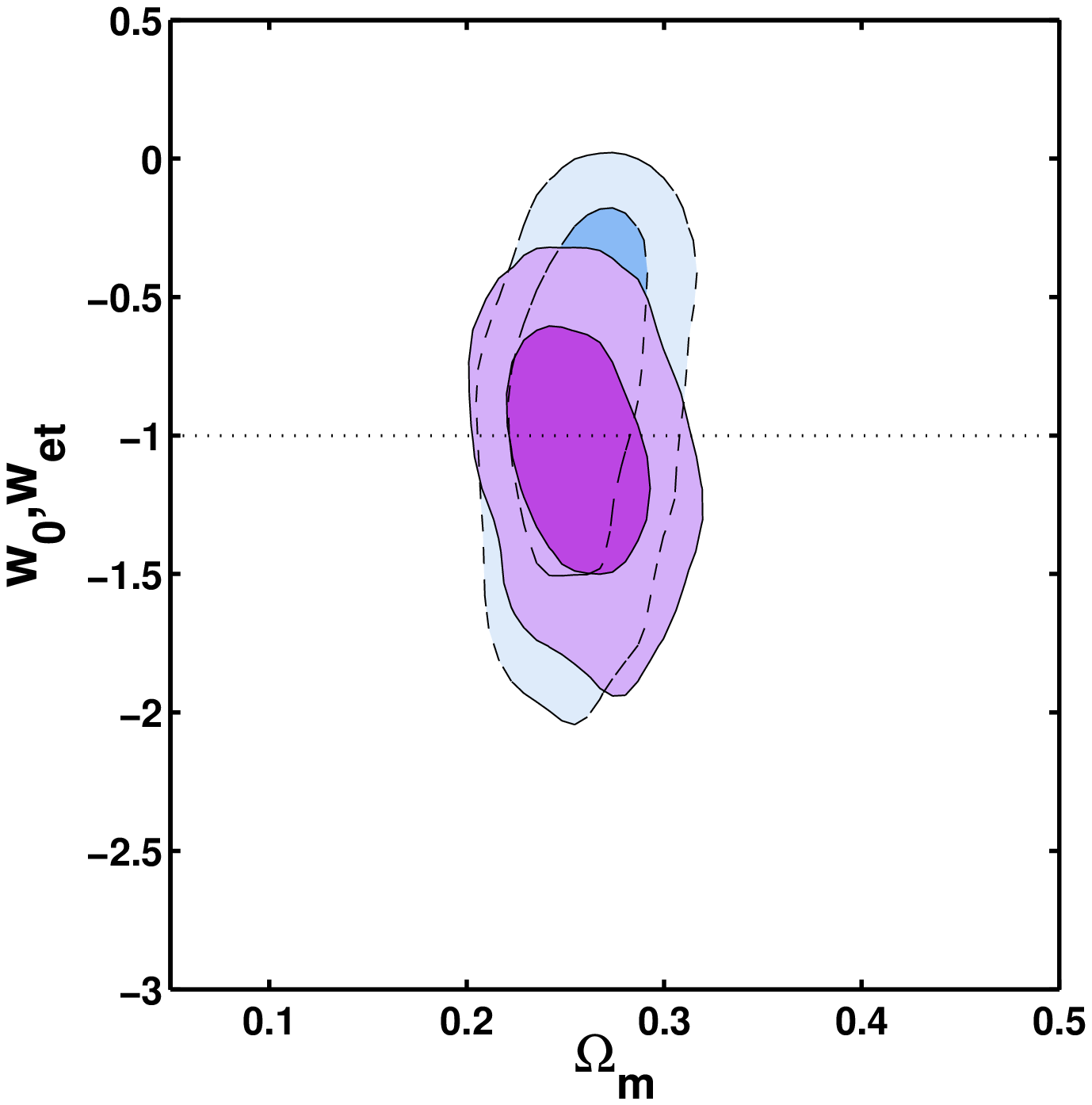,width=0.43 \textwidth,angle=0}
\hspace{1.3cm}\psfig{figure=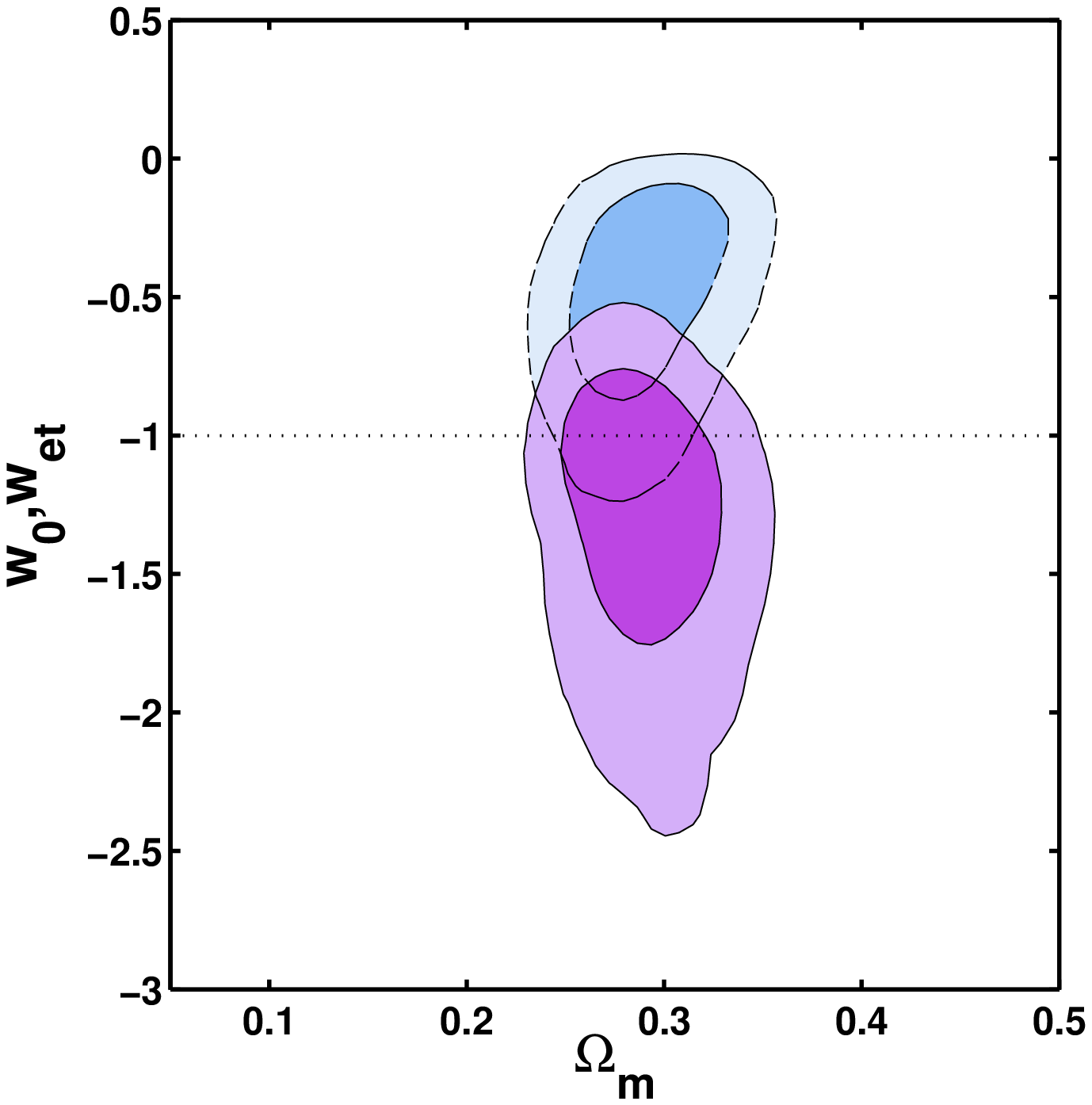,width=0.43 \textwidth,angle=0}
}
\caption{The 68.3 and 95.4 per cent confidence limits in the
($\Omega_{\rm m}$;$w_{\rm 0}$,$w_{\rm et}$) plane determined from the
\fgas+CMB+SNIa data using our most general dark energy model
(Equation~\ref{eq:evolve}) with the transition scale factor
marginalized over the range $0.5< a_{\rm t}< 0.95$. The solid, purple contours
show the results on ($\Omega_{\rm m}$,$w_{\rm 0}$). The dashed, turquoise lines
show the results on ($\Omega_{\rm m}$,$w_{\rm et}$). The horizontal
dotted line denotes the cosmological constant model ($w_{\rm 0}=w_{\rm
et}=-1$). The left and right panels show the results obtained for the
two SNIa samples: (Left panel) Davis \etal (2007) and (Right panel)
Riess \etal (2007). A flat geometry ($\OK=0$) is assumed. 
The data provide no significant evidence for
evolution in $w$ and are consistent with the cosmological
constant ($\Lambda$CDM) model ($w=-1$; Section~\ref{section:evol}).
}\label{fig:w_evol}
\end{figure*}

Fig.~\ref{fig:w_evol} shows the constraints on $w_0$ and $w_{\rm et}$
obtained from a combined analysis of \fgas+CMB+SNIa data using the
general, evolving dark energy model (Equation~\ref{eq:evolve})
and assuming geometric flatness ($\OK=0$). The
left and right panels show the results obtained for the two separate
SNIa samples (Section~\ref{section:otherdata}). Using the Davis \etal
(2007) SNIa compilation (left panel), we find no evidence for
evolution in the dark energy equation of state over the redshift range
spanned by the data: the results on the dark energy equation of state
at late and early times, $w_0=-1.05^{+0.31}_{-0.26}$ and $w_{\rm
et}=-0.83^{+0.48}_{-0.43}$ (68 per cent confidence limits), are both
consistent with a cosmological constant model ($w=-1$, constant).  A
similar conclusion is drawn by Davis \etal (2007) using 
SNIa+CMB+Baryon Acoustic Oscillation (BAO) data.

We note, however, a hint of evolution in the dark energy equation of
state when the Riess \etal (2007) `gold' SNIa sample is used instead
(right panel of Fig.~\ref{fig:w_evol}). In this case, the marginalized
constraints on dark energy at late and early times, as defined in
Section~\ref{section:demod}, differ at the $2-3\sigma$ level.  Similar
indications are also apparent in the analysis of the same SNIa
(+CMB+BAO) data by Riess \etal (2007).  However, the analysis using
the Davis \etal (2007) SNIa compilation (left panel), which includes
the high-quality, high-redshift HST supernovae from Riess \etal (2007)
and which shows no suggestion of a departure from the $\Lambda$CDM
model, argues that the hint of evolution in the right panel of
Fig.~\ref{fig:w_evol} may be systematic in origin (see also Riess
\etal 2007 and Conley \etal 2007 for discussions).

\subsection{The degeneracy breaking power of the combined 
\fgas+CMB(+SNIa) data}
\label{section:degenbreak}

\begin{figure*}
\vspace{0.2cm}
\hbox{
\hspace{0.2cm}\psfig{figure=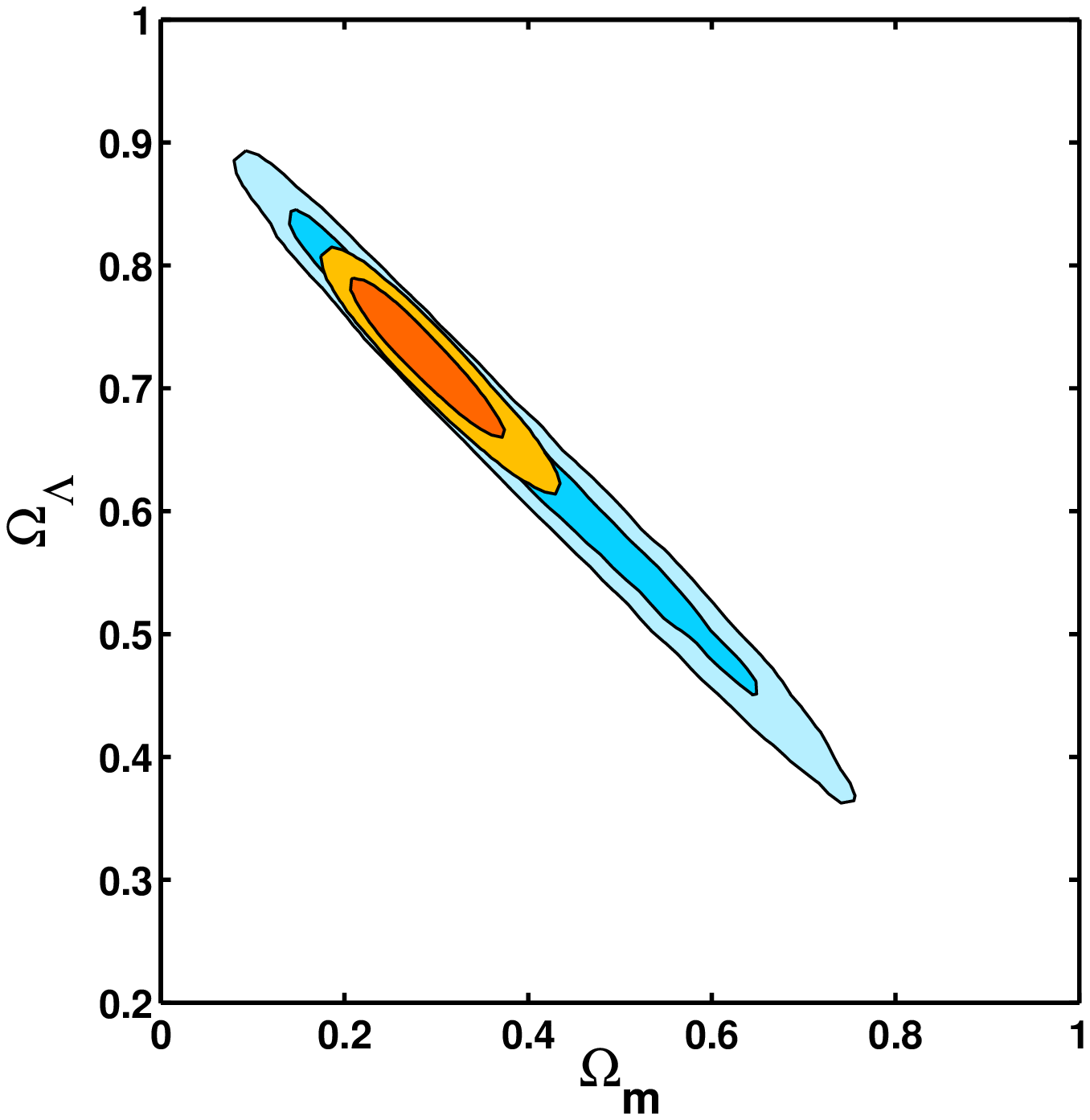,width=0.43 \textwidth,angle=0}
\hspace{1.3cm}\psfig{figure=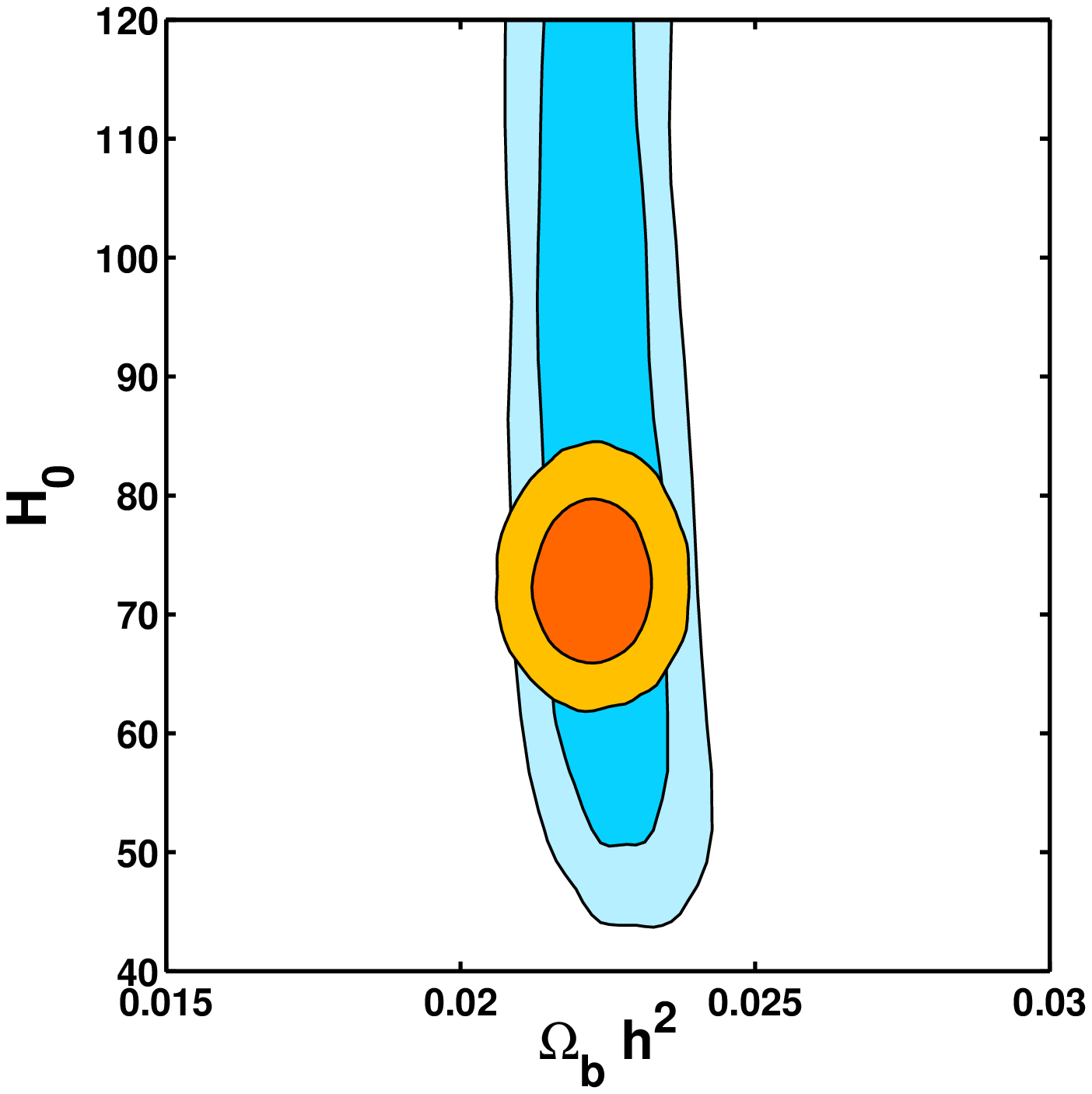,width=0.43 \textwidth,angle=0}
}
\caption{The degeneracy-breaking power of the \fgas+CMB data. 
Contours show the 68.3 and 95.4 per cent confidence limits
determined from the CMB data alone (larger, blue contours) and 
combined \fgas+CMB data (smaller, orange contours). (Left panel) the 
constraints on $\OM$ and $\ODE$ for the $\Lambda$CDM model
with the curvature included as a free parameter. (Right panel)
The tight constraints on $H_0$ and $\OBH$ for the flat, constant $w$
model, demonstrating why external priors on these  
two parameters are not required when the \fgas~and CMB data are combined. 
}\label{fig:fgascmb_degen}
\end{figure*}

The degeneracy breaking power of the combined \fgas+CMB data set is
evidenced in the left panel of Fig.~\ref{fig:fgascmb_degen}, which
shows the constraints on $\OM$ versus $\ODE$ for a $\Lambda$CDM model
with free curvature for the CMB data alone (blue contours) and the
combined \fgas+CMB data set (orange contours). For the \fgas+CMB data,
we measure $\Omega_{\rm m}=0.278^{+0.064}_{-0.050}$ and
$\Omega_{\Lambda}=0.732^{+0.040}_{-0.046}$ (68 per cent confidence
limits), with the curvature $\OK=-0.011^{+0.015}_{-0.017}$. As
mentioned above, no external priors on $\OBH$ and $h$ are required
when the \fgas~and CMB data are combined. The degeneracy breaking 
power of other combinations of data with the CMB is discussed by
Spergel \etal (2007).

The right panel of Fig.~\ref{fig:fgascmb_degen} shows the constraints
on the Hubble Constant, $H_0$, and mean baryon density, $\OBH$,
determined using the flat, constant $w$ model for the CMB data alone
(blue contours) and the combined \fgas+CMB data set (orange
contours). The improvement in the constraints on these parameters
determined from the \fgas+CMB data over the CMB data alone is
substantial. The tight constraints for the \fgas+CMB data,
$H_0=72.5\pm4.6$\kmpspMpc~and $\OBH=0.0223\pm0.0007$, demonstrate
clearly why external priors on these two parameters are not required
when the \fgas~and CMB data are combined.  Indeed, the constraints on
$H_0$ and $\OBH$ obtained from the \fgas+CMB data are significantly
tighter than the external priors on these parameters that are employed
when the \fgas~data are used alone (Table~\ref{table:sys}). Similar
constraints on $H_0$ and $\OBH$ are presented by the WMAP team
(Spergel \etal 2007) for flat $\Lambda$CDM models using various data
combinations.

\begin{figure*}
\vspace{0.2cm}
\hbox{
\hspace{0.2cm}\psfig{figure=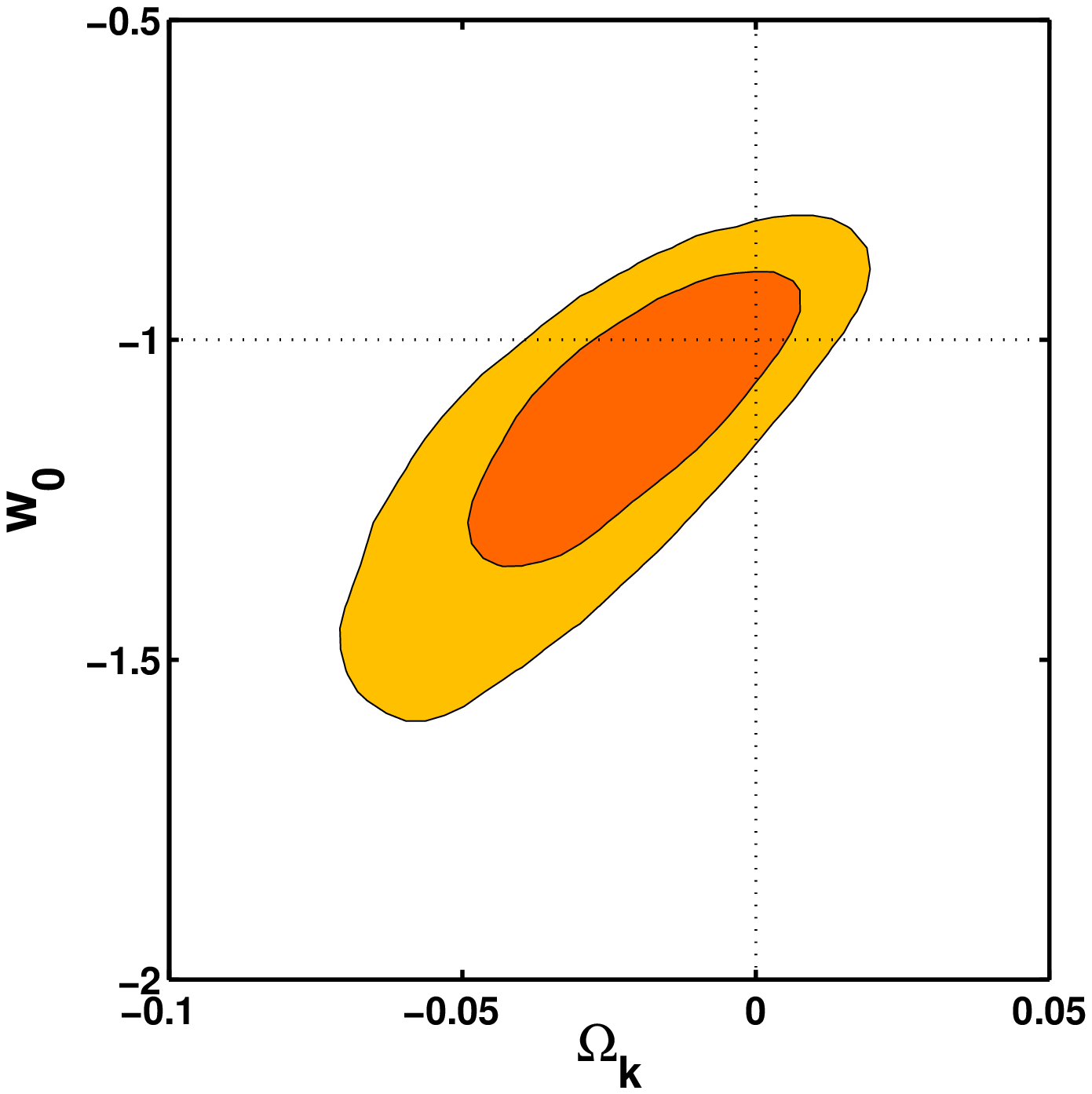,width=0.43 \textwidth,angle=0}
\hspace{1.3cm}\psfig{figure=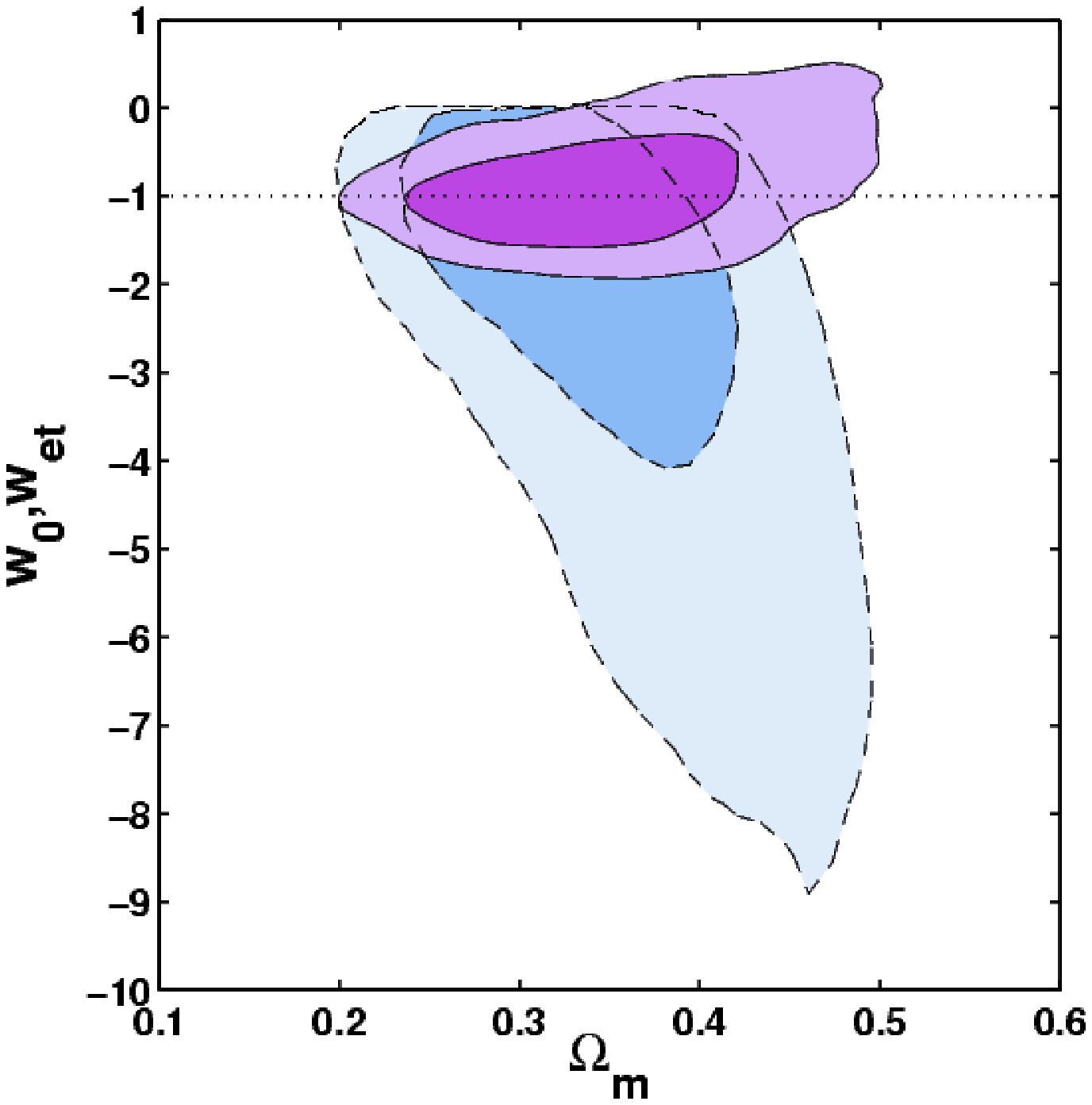,width=0.43 \textwidth,angle=0}
}
\caption{(Left panel) The 68.3 and 95.4 per cent confidence limits
on the dark energy equation of state and curvature from the analysis 
of the \fgas+CMB+SNIa data using the non-flat, constant $w$ model. The SNIa
compilation of Davis \etal (2007) has been used. The horizontal and 
vertical dotted lines denote the loci for cosmological constant 
models and geometric flatness, respectively, both of which are
consistent with the data. (Right panel) 
The 68.3 and 95.4 per cent confidence limits in the
($\Omega_{\rm m}$;$w_{\rm 0}$,$w_{\rm et}$) plane 
determined from the \fgas+CMB+SNIa data for the general dark energy model
(Equation~\ref{eq:evolve}) with the curvature also included
as a free parameter. Other details as in the left panel of 
Fig.~\ref{fig:w_evol}.
}\label{fig:w_nonflat}
\end{figure*}

Fig.~\ref{fig:w_nonflat} shows the constraints on the dark energy
equation of state obtained from an analysis of the combined
\fgas+CMB+SNIa data set where the curvature is also included as a free
parameter.  The marginalized results for the constant $w$ model (left
panel), $w=-1.08^{+0.13}_{-0.19}$ and $\OK=-0.024^{+0.022}_{-0.018}$,
are comparable to those of Spergel \etal (2007; see their Fig. 17)
from a combined analysis of CMB, SNIa and galaxy redshift survey
data.  The constraints for the non-flat evolving $w$ model (right
panel), though weaker than those for the flat model
(Fig.~\ref{fig:w_evol}), remain interesting and are also consistent
with a cosmological constant.  As discussed by Rapetti \etal (2005;
see also Spergel \etal 2007), such results demonstrate the power of
the \fgas+CMB+SNIa data to constrain the properties of dark energy
without the need to assume that the Universe is flat.

Using the non-flat evolving $w$ model but fixing the transition
redshift $z_{\rm t}=1$ in Equation~\ref{eq:evolve}, we recover the
model used by the Dark Energy Task Force (DETF) to assess the power of
future dark energy experiments. The combination of current
\fgas+CMB+SNIa data provides a DETF figure of merit $\sim 2$.

\section{Discussion}

The new Chandra \fgas~results and analysis presented here build upon
those of Allen \etal (2004) and Rapetti \etal (2005). The present
study includes 16 more objects, approximately twice as much Chandra
data and extends the study beyond a redshift of 1. Our analysis
includes a comprehensive and conservative treatment of systematic
uncertainties (Section~\ref{section:modelling}; see also
Table~\ref{table:sys}). Allowances for such uncertainties are easily
incorporated into the MCMC analysis.

As with SNIa studies, the \fgas~data constrain dark energy via its
effects on the distance-redshift relation to a well-defined source
population -- in this case, the largest, dynamically relaxed galaxy
clusters -- using measurements of a `standard' astrophysical quantity
-- the ratio of baryonic-to-total mass in the clusters. Our results
provide a clear and independent detection of the effects of dark
energy on the expansion of the Universe at $\sim 99.99\%$ confidence
for a standard non-flat $\Lambda$CDM model, an accuracy comparable to
that obtained from current SNIa work (\eg Astier \etal 2006; Riess
\etal 2007; Wood-Vasey \etal 2007; Miknaitis \etal 2007).  Like SNIa
studies, the \fgas~data trace the evolution of dark energy over the
redshift range $0<z<1$, where it grows to dominate the overall energy
density of the Universe. Our results for the \fgas~data alone, and the
combination of \fgas+CMB+SNIa data, show that this growth is
consistent with that expected for models in which the dark energy is a
cosmological constant ($w=-1$).

Despite some clear similarities, important complementary differences
between the \fgas~and SNIa experiments exist. In the first case,
the physics of the astrophysical objects -- large, relaxed galaxy
clusters and SNIa -- are very different; the fact that such similar
cosmological results are obtained from the distance-redshift
information for these separate source populations is 
reassuring. Future studies, combining the two techniques but using
larger target samples, should open the possibility for precise
distance-redshift measurements and good control of systematic
uncertainties, employing both kinematic and dynamical analyses (\eg
Rapetti \etal 2007; Riess \etal 2007 and references therein).

An important strength of the \fgas~method is the tight constraint on
$\OM$ provided by the normalization of the \fgas~curve; this breaks
the degeneracy between the mean matter density and dark energy density
inherent in the distance measurements. Our result on $\OM$ is
consistent with a host of previous X-ray studies (Section 1).  

A further strength, which is of relevance when
considering observing strategies for future dark energy work, is the
small intrinsic dispersion in the \fgas~distance measurements. SNIa
studies have established the presence of a systematic scatter of $\sim
7$ per cent in distance measurements for individual SNIa using high
quality data (Jha \etal 2007; see also \eg Riess \etal 2004, 2007;
Astier \etal 2006; Wood-Vasey \etal 2007).  In contrast, systematic
scatter remains undetected in the present Chandra \fgas~data for hot,
relaxed clusters, despite the fact that the weighted mean
$statistical$ scatter in $f_{\rm gas}$ data corresponds to only $\sim
5$ per cent in distance. This small systematic scatter for large,
dynamically relaxed clusters (identified as relaxed on the basis of
their X-ray morphologies) is consistent with the predictions from
hydrodynamical simulations (\eg Nagai \etal 2007a), although the results
for both observed and simulated clusters are, at present, based on relatively
small samples and more data are required.  We stress that such small
systematic scatter is neither expected nor observed in studies where a
restriction to morphologically relaxed clusters is $not$ employed \eg
compare the small scatter measured here with the much larger scatter
observed in the studies of LaRoque \etal (2006) and Ettori \etal
(2003); see also Nagai \etal (2007a). The restriction to the hottest,
relaxed clusters, for which \fgas~is independent of temperature
(Fig.~\ref{fig:fgaskt}), also simplifies the determination of
cosmological parameters.

As mentioned above, the allowances for systematic uncertainties
included in the analysis are relatively conservative. Much progress is
expected over the coming years in refining the ranges of these
allowances, both observationally and through improved simulations. As
discussed in Sections 5.1 and 5.2, a reduction in the size of the
required systematic allowances will tighten the cosmological
constraints. Improved numerical simulations of large samples of
massive clusters, including a more complete treatment of star
formation and feedback physics that reproduces both the observed
optical galaxy luminosity function and cluster X-ray properties, will
be of major importance. Progress in this area has been made
(\eg Bialek, Evrard \& Mohr 2001, Muanwong \etal 2002, Kay S. \etal,
2004, Kravtsov, Nagai \& Vikhlinin 2005, Ettori \etal 2004, 2006,
Rasia \etal 2006; Nagai \etal 2007a,b), though more work remains.  In
particular, this work should improve the predictions for $b(z)$.
Further deep X-ray and optical observations of nearby clusters will
provide better constraints on the viscosity of the cluster gas.
Improved optical/near infrared observations of clusters should pin
down the stellar mass fraction in galaxy clusters and its evolution.

Ground and space-based gravitational lensing studies will provide
important, independent constraints on the mass distributions in
clusters; a large program using the Subaru telescope
and Hubble Space Telescope is underway, as is similar work by
other groups (\eg Hoekstra 2007). Follow-up observations of the SZ
effect will also provide additional, independent constraining power in
the measurement of cosmological parameters (the combination of direct
observations of the SZ effect using radio/sub-mm data and the
prediction of this effect from X-ray data provides an additional
constraint on absolute distances to the clusters \eg Molnar \etal
2002, Schmidt, Allen \& Fabian 2004; Bonamente \etal 2006 and
references therein). Moreover, the independent constraints provided by
the SZ observations should allow a reduction of
the priors required in future work (\eg Rapetti \&
Allen 2007).

In the near future, continuing programs of Chandra and XMM-Newton
observations of known, X-ray luminous clusters should allow important
progress to be made, both by expanding the \fgas~sample (\eg Chandra
snapshot observations of the entire MACS sample; Ebeling \etal 2001,
2007) and through deeper observations of the current target list. The
advent of new, large area SZ surveys (\eg Ruhl \etal 2004) will soon
provide important new target lists of hot, X-ray luminous high
redshift clusters. A new, large area X-ray survey such as that
proposed by the Spectrum-RG/eROSITA
project\footnote{http://www.mpe.mpg.de/projects.html\#erosita} could
make a substantial contribution, finding hundreds of suitable systems
at high redshifts.

Looking a decade ahead, the Constellation-X Observatory
(Con-X)\footnote{http://constellation.gsfc.nasa.gov/} and, later,
XEUS\footnote{http://www.rssd.esa.int/index.php?project=XEUS} offer
the possibility to carry out precise studies of dark energy using the
\fgas~technique.  As discussed by Rapetti \& Allen (2007; see also
Rapetti \etal 2006), the large collecting area and combined
spatial/spectral resolving power of Con-X should permit precise
\fgas~measurements with $\sim 5$ per cent accuracy for large samples
($\approxgt 500$) of hot, massive clusters ($kT\approxgt 5$keV)
spanning the redshift range $0<z<2$ (typical redshift $z\sim0.6$).
The predicted constraints on dark energy from such an experiment,
assuming Planck priors (Albrecht \etal 2006), have a DETF figure of
merit $\approxgt 20$, which is comparable to other leading proposed
dark energy techniques such as SNIa, cluster number counts, weak
lensing and baryon acoustic oscillation studies. The high spectral
resolution offered by the Con-X calorimeters will also permit precise
measurements of bulk motions and viscosity in the cluster gas,
addressing directly one of the main sources of systematic uncertainty
in the method.

An ASCII table containing the redshift and $f_{\rm gas}(z)$ data is
available at {\it http://xoc.stanford.edu} or from the authors on
request. The analysis code, in the form of a patch to CosmoMC, will be
made available at a later date.

\section*{Acknowledgements}

We thank Sarah Church, Vince Eke, Bob Kirshner, Gary Mamon, Herman Marshall, 
Rich Mushotzky, Jerry Ostriker, Harvey Tananbaum, Alexey Vikhlinin, 
Jochen Weller and Nick White for discussions over the course of this work. 
We also thank Antony Lewis for help with CosmoMC. The
computational analysis was carried out using the KIPAC
XOC compute cluster at the Stanford Linear Accelerator Center
(SLAC). We acknowledge support from the National Aeronautics and Space
Administration through Chandra Award Numbers DD5-6031X, GO2-3168X,
GO2-3157X, GO3-4164X, GO3-4157X and G07-8125X, issued by the Chandra X-ray
Observatory Center, which is operated by the Smithsonian Astrophysical
Observatory for and on behalf of the National Aeronautics and Space  
Administration under contract NAS8-03060. This work was supported in
part by the U.S. Department of Energy under contract number
DE-AC02-76SF00515.

\end{document}